\documentclass[aps,prl,twocolumn,superscriptaddress]{revtex4-1}

\usepackage{amsfonts}
\usepackage{amssymb}
\usepackage{amsmath}
\usepackage{upgreek}
\usepackage{color}

\usepackage{graphicx}
\usepackage{xr-hyper}
\usepackage[colorlinks=true,citecolor=blue,linkcolor=blue,urlcolor=blue,filecolor=blue]{hyperref}

\makeatletter
\newcommand*{\addFileDependency}[1]{
  \typeout{(#1)}
  \@addtofilelist{#1}
  \IfFileExists{#1}{}{\typeout{No file #1.}}
}
\makeatother

\newcommand{\UIUC}{Department of Physics and Materials Research Laboratory, University of Illinois, Urbana, Illinois 61801, USA}
\newcommand{\Amsterdam}{Institute for Theoretical Physics Amsterdam, University of Amsterdam, Science Park 904, 1098 XH Amsterdam, The Netherlands}
\newcommand{\Bristol}{School of Physics, Tyndall Avenue, Bristol BS8 1TL, United Kingdom}
\newcommand{\Donostia}{Donostia International Physics Center, P. Manuel de Lardizabal 4, 20018 Donostia-San Sebasti\'an, Spain}
\newcommand{\UCLA}{Department of Physics and Astronomy, University of California at Los Angeles, Los Angeles, California 90095, USA}
\newcommand{\PKU}{International Center for Quantum Materials, School of Physics, Peking University, Beijing 100871, China}
\newcommand{\APS}{Materials Science Division, Argonne National Laboratory, Lemont, Illinois 60439, USA}
\newcommand{\UMD}{Maryland Quantum Materials Center, Department of Physics, University of Maryland, College Park, Maryland 20742, USA}
\newcommand{\DU}{Department of Physics, Drexel University, Philadelphia, Pennsylvania 19104, USA}
\newcommand{\CHESS}{CHESS, Cornell University, Ithaca, New York 14853, USA}
\newcommand{\CIFAR}{Canadian Institute for Advanced Research, Toronto, Ontario M5G 1Z8, Canada}
\newcommand{\IKERBASQUE}{IKERBASQUE, Basque Foundation for Science, Plaza Euskadi 5, 48009 Bilbao, Spain}
\newcommand{\UMN}{School of Physics and Astronomy, University of Minnesota, Minneapolis, Minnesota 55455, USA}

\newcommand{\mytitle}{In-plane anisotropy of charge density wave fluctuations in 1$T$-TiSe$_2$}

\newcommand{\dateOfSubmission}{\today}

\begin{document}

\title{\mytitle}

\author{Xuefei Guo}
\email{xuefeig2@illinois.edu}
\affiliation{\UIUC}
\author{Anshul Kogar}
\email{anshulkogar@physics.ucla.edu}
\affiliation{\UCLA}
\author{Jans Henke}
\affiliation{\Amsterdam}
\author {Felix Flicker}
\affiliation{\Bristol}
\author {Fernando de Juan}
\affiliation{\Donostia}
\affiliation{\IKERBASQUE}
\author{Stella X.-L. Sun}
\affiliation{\UIUC}
\author{Issam Khayr}
\affiliation{\UIUC}
\affiliation{\UMN}
\author{Yingying Peng}
\affiliation{\PKU}
\author{Sangjun Lee}
\affiliation{\UIUC}
\author{Matthew J. Krogstad}
\author{Stephan Rosenkranz}
\author{Raymond Osborn}
\affiliation{\APS}
\author{Jacob P. C. Ruff}
\affiliation{\CHESS}
\author{David B. Lioi}
\author{Goran Karapetrov}
\affiliation{\DU}
\author{Daniel J. Campbell}
\affiliation{\UMD}
\author{Johnpierre Paglione}
\affiliation{\UMD}
\affiliation{\CIFAR}
\author{Jasper van Wezel}
\affiliation{\Amsterdam}
\author{Tai C. Chiang}
\author{Peter Abbamonte}
\email{abbamont@illinois.edu}
\affiliation{\UIUC}

\date{\dateOfSubmission}

\begin{abstract}
We report measurements of anisotropic triple-$q$ charge density wave (CDW) fluctuations in the transition metal dichalcogenide 1$T$-TiSe$_2$ over a large volume of reciprocal space with X-ray diffuse scattering. Above the transition temperature, $T_{\text{CDW}}$, the out-of-plane diffuse scattering is characterized by rod-like structures which indicate that the CDW fluctuations in neighboring layers are largely decoupled. In addition, the in-plane diffuse scattering is marked by ellipses which reveal that the in-plane fluctuations are anisotropic. 
Our analysis of the diffuse scattering line shapes and orientations suggests that the three charge density wave components contain independent phase fluctuations. At $T_{\text{CDW}}$, long range coherence is established in both the in-plane and out-of-plane directions, consistent with the large observed value of the CDW gap compared to $T_{\text{CDW}}$, and the predicted presence of a hierarchy of energy scales.

\end{abstract}

\maketitle

Fluctuations are central to the critical behavior observed near continuous phase transitions~\cite{goldenfeld2018lectures, collins1989magnetic}. In quasi-two-dimensional layered systems, where inter-plane coupling is weak, fluctuations can persist to temperatures far exceeding the transition temperature and can affect the system's normal state electronic properties. In these cases, in-plane fluctuations start to build below the mean field transition temperature, $T_\text{MF}$, but they initially remain uncorrelated between layers. As the temperature is reduced further and approaches the critical temperature, $T_{\text{c}}$, a 2D-to-3D crossover occurs as the inter-plane correlation length starts to exceed the inter-plane distance. At $T_{\text{c}}$, both in-plane and out-of-plane correlation lengths diverge and full three-dimensional long range order sets in \cite{collins1989magnetic}. A wide variety of physical systems have been shown to conform to this simple description~\cite{collins1989magnetic, Donath2008, hirakawa1982kosterlitz, ron2019dimensional, valla2002coherence, Wildes2006, Wildes2015}; because transition metal dichalcogenides are layered compounds, and their charge density wave (CDW) order is characterized by large ratios of $2\Delta(0)/k_BT_{\text{CDW}}$, where $\Delta(0)$ is the zero-temperature gap and $T_{\text{CDW}}$ is the charge density wave transition temperature, fluctuations in these materials should abide by a similar phenomenology. 

\begin{figure}
    \centering
    \includegraphics[width=\columnwidth]{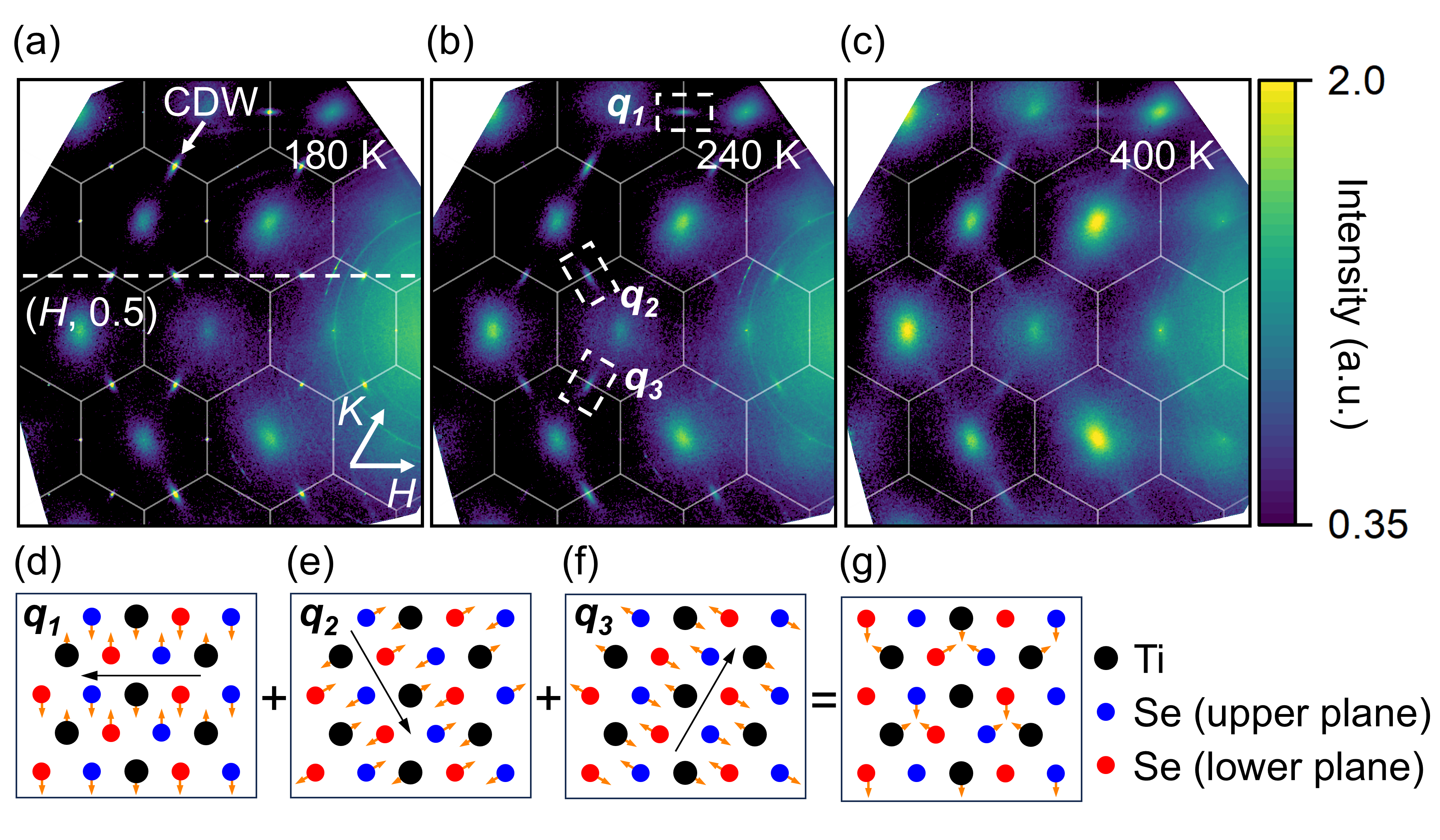}
    \caption{In-plane anisotropy of CDW fluctuations. (a)-(c) In-plane momentum maps at $L = 0.5$ r.l.u. for temperatures below (180\,K), above (240\,K), and well above (400\,K) $T_{\text{CDW}}$($= 195$\,K). Below $T_{\text{CDW}}$, sharp peaks are observed at the Brillouin zone boundaries (white line), indicating the presence of an ordered state. As the CDW is melted, CDW fluctuations become elongated along the in-plane CDW vectors $q_i$ ($i= 1, 2, 3)$. The diffuse feature at the center of the Brillouin zone corresponds to the tail of the structural Bragg peaks from neighboring $L-$planes. The intensity scale bar is in arbitrary units and follows a logarithmic scale. (d)-(f) Schematic depictions of the in-plane displacement pattern of the three components contributing to the triple-$q$ CDW. (d) $q_1$-type component, (e) $q_2$-type, and (f) $q_3$-type. Black arrows indicate the direction of the propagation vector for each CDW component, while small orange vectors indicate the atomic displacement directions that would result from having only each single component present. (g) The full in-plane triple-$q$ distortion pattern obtained by adding the atomic displacements from all three CDW components. Black circles represent Ti atoms, blue circles represent Se atoms in the upper plane and red circles represent Se atoms in the lower plane.}
    \label{fig1}
\end{figure}

This picture is complicated, however, in the CDW-hosting transition metal dichalcogenides (TMDs) by the fact that their charge order consists of three unidirectional components, in a structure that is often referred to as a triple-$q$ CDW (see Fig.~\ref{fig1}(d)-(g))~\cite{wilson1975charge}. The triple-$q$ structure preserves the threefold crystallographic in-plane symmetry, despite the unidirectionality of the individual CDW components~\cite{rossnagel2011origin}. Preservation of this symmetry gives an energetic advantage to triple-$q$ order over CDWs containing only a single component in these compounds, which can be described by an effective ``attraction'' between CDW components in their free energy description (see Supplemental Material)~\cite{mcmillan1975}. For the same reason, short-range ordered regions above $T_{\text{CDW}}$ will consist of local triple-$q$ order, while the large value of 2$\Delta(0)$/k$_B T_{\text{CDW}} \approx 8.7$~\cite{Li2007} implies that long-range phase coherence is prevented at those temperatures by the proliferation of phase fluctuations rather than the suppression of the CDW amplitude~\cite{NbSe2_1,NbSe2_2}. Phase slips may occur either in all three components simultaneously, or independently in different components. As described in the Supplemental Material, the latter option is generically favored in the free energy description for a triple-$q$ CDW.

\begin{figure}
    \centering
    \includegraphics[width=\columnwidth]{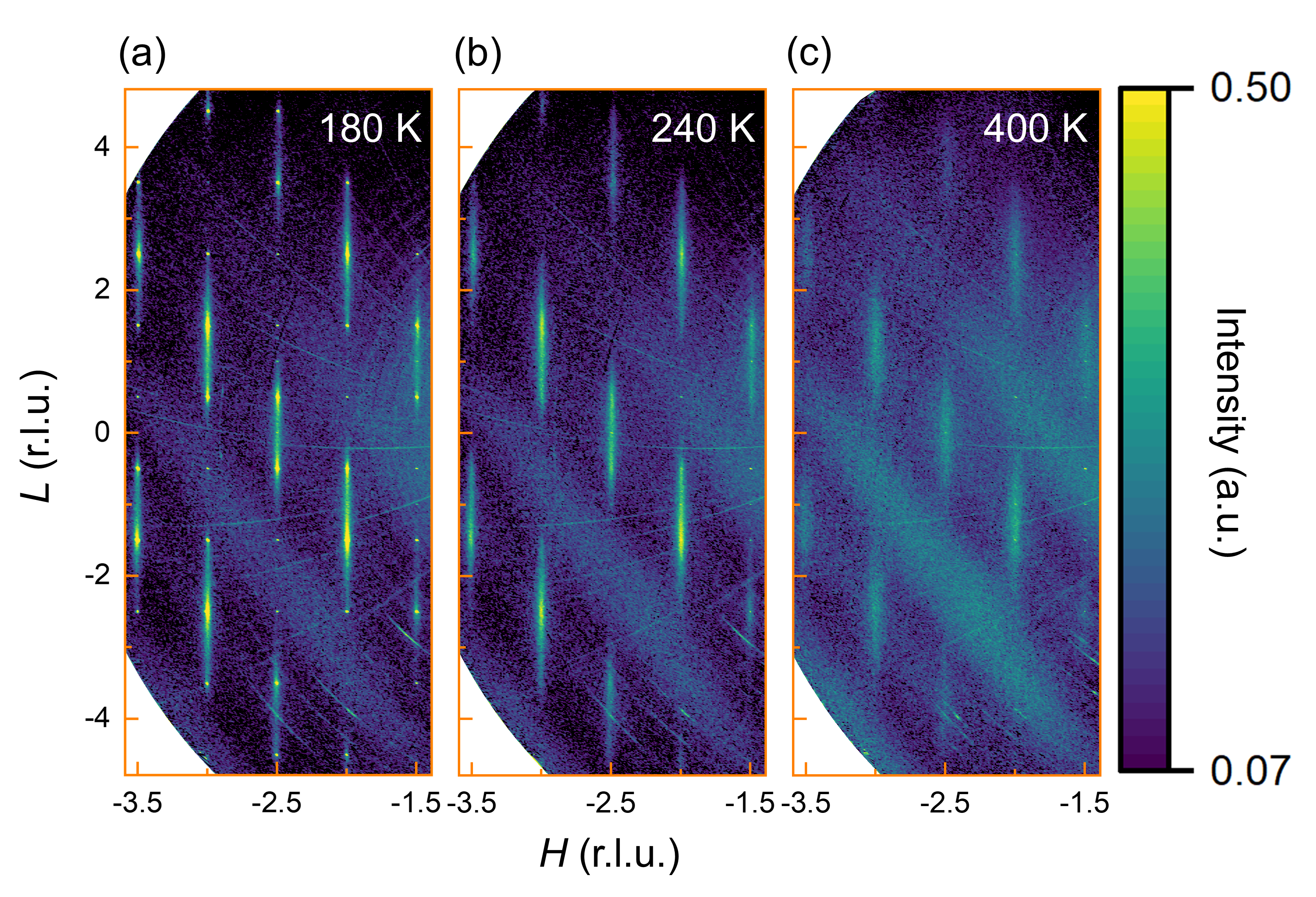}
    \caption{($H$, $L$) momentum maps at $K = 0.5$ r.l.u. showing sharp CDW diffraction peaks and diffuse scattering arising from CDW fluctuations. (a) At 180\,K ($\textless T_{\text{CDW}} (=195$\,K)), the observed pattern shows sharp peaks, indicative of long range CDW order. (b) Above $T_{\text{CDW}}$, at 240\,K, only diffuse scattering in the form of a `raindrop'-like pattern is clearly visible which indicates the presence of significant inter-layer phase fluctuations (CDW stacking faults). (c) Well above $T_{\text{CDW}}$ at 400\,K, structures within the `raindrops' have disappeared and featureless `rods' remain, which indicates the loss of all out-of-plane phase coherence. The intensity modulation with roughly period four visible in all panels is due to a geometric structure factor. The $K = 0.5$ r.l.u. momentum cut is shown schematically in Fig.~\ref{fig1}(a) with respect to the in-plane elliptical diffuse scattering. The scale bar is in arbitrary units and follows a logarithmic scale.}
    \label{fig2}
\end{figure}

In the usual 2D-to-3D crossover scenario, in-plane fluctuations form regions that are, on average, quasi-two-dimensional and isotropic to within the crystalline symmetry~\cite{collins1989magnetic}. Because the CDW order in 1$T$-TiSe$_2$ contains three components with independent phase fluctuations, however, both the fluctuating regions and the crossover to three-dimensional order may obtain a more structured character. In this Letter, we show that while out-of-plane diffuse scattering indicates decoupled CDW fluctuations in neighboring layers, the in-plane diffuse scattering arising from individual CDW components is anisotropic, with a directionally-dependent characteristic length scale. We argue that, in the temperature range near but above $T_{\text{CDW}}$, each CDW component contains independent phase slips. While the combined triple-$q$ short-ranged order remains isotropic on average, the mean distance between phase slips for any given component is anisotropic and determines its distinct correlation length scales in different directions. We thus demonstrate that despite its threefold symmetry and quasi-two-dimensional crystal structure, 1$T$-TiSe$_2$ possesses anisotropic CDW fluctuations in the basal plane and a hierarchy of three distinct length scales arising from independently fluctuating CDW-components~\cite{vanWezel2010}.

To carry out our experiments, we used two different batches of TiSe$_2$ single crystals with different synthesis methods. We synthesised vacancy-reduced TiSe$_2$ crystals using a Se-flux method under high pressure \cite{Campbell2019}. These are insulating at low temperatures; their resistivity is hysteretic from 30-80 K. We grew another batch of samples using iodine vapour transport \cite{iavarone2012evolution, husanikova2013magnetization}. These are crystallographically cleaner with less mosaicity; they are semimetallic at low temperatures with higher carrier concentrations. Both batches of samples give identical diffuse scattering results. The data presented in the main manuscript are from the semimetallic sample, while the data for the vacancy-reduced sample are provided in the Supplemental Material. 

We performed hard x-ray diffraction measurements at Cornell High Energy Synchrotron Source using a photon energy of 27.3\,keV. High energy data at 56.7\,keV and additional data from the Advanced Photon Source with a photon energy of 87.4 keV are shown in the Supplemental Material. We used a transmission geometry for these measurements with a single-photon counting detector suited for high-energy diffraction. We glued the samples onto Kapton capillaries and cooled them by a helium or nitrogen gas jet. At each temperature, we swept the sample through a full rotation to obtain three dimensional momentum mappings \cite{krogstad2020reciprocal}. 

1$T$-TiSe$_2$ is a layered van der Waals material with an octahedrally coordinated structure \cite{rossnagel2011origin}. Upon cooling below the transition temperature ($T_{\text{CDW}}=195$\,K in the semimetallic sample), 1$T$-TiSe$_2$ undergoes a second-order phase transition into a commensurate 2$\times$2$\times$2 CDW superstructure. The microscopic driving mechanism giving rise to the CDW order has been controversial for decades, though the emerging consensus is that both excitonic and electron-phonon effects contribute~\cite{Kogar2017, vanWezel2010, Cercellier2007, porer2014non, Burian2021, rossnagel2011origin}. Regardless, triple-$q$ long-range charge order forms, with the three CDW wave vectors pointing 120$^{\circ}$ apart when projected onto the plane (Fig.~\ref{fig1}(d)-(g)).

In Fig.~\ref{fig1}(a) we show that CDW satellite peaks form below $T_{\text{CDW}}=195$K. Three distinct CDW components $q_1$, $q_2$, and $q_3$ are observed at the boundaries of the normal state Brillouin zone (labeled in Fig.~\ref{fig1}(b)). Warming the sample above the transition temperature to 240~K, diffuse scattering near the CDW wave vectors is the dominant feature, as reported previously~\cite{holt2001x}. The diffuse scattering intensity is at least an order of magnitude weaker than the CDW peak intensities at 180~K. As can clearly be seen in Fig.~\ref{fig1}(b) and (c), the diffuse scattering is anisotropic in the plane. For each of the three CDW peaks, the widths along the major axis of the ellipse are roughly three times those along the minor axis at 240~K. These ellipses effectively become streaks at 400~K, the highest temperature measured in our study (Fig.~\ref{fig1}(c)). Importantly, the ellipses for each scattering peak form with their major axes in the direction parallel to the corresponding in-plane CDW wave vector. The orientation of these ellipses implies that the Ti and Se atoms involved in individual CDW components form chain-like structures transverse to each in-plane CDW wave vector, with inter-chain coupling weaker than the intra-chain coupling (Fig.~\ref{fig1}(d)-(g)). These observations are consistent with the proposed hierarchy of energy scales in 1$T$-TiSe$_2$~\cite{vanWezel2010}. 
\begin{figure}
    \centering
    \includegraphics[width=\columnwidth]{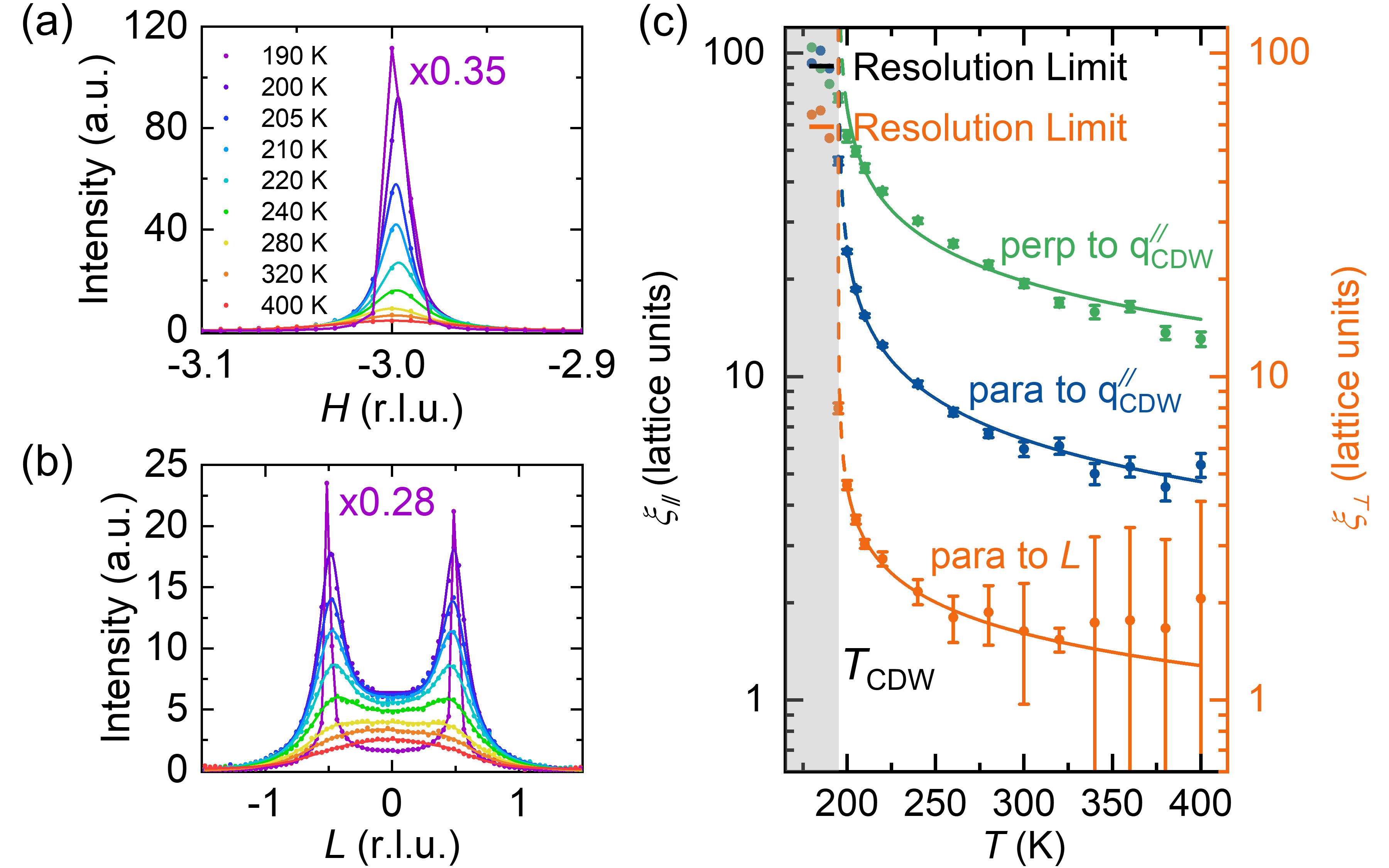}
    \caption{Momentum scans and correlation lengths at different temperatures. (a) In-plane momentum scans for (-3.0, 1.5, 0.5) along $H$ below and above $T_{\text{CDW}}(=$195\,K). The constant background far away from the CDW peak is subtracted. Below $T_{\text{CDW}}$, the resolution-limited sharp peak indicates the presence of long range order, with the line serving as a guide to the eye. Above $T_{\text{CDW}}$, the peaks are fitted with Lorentzians, indicating fluctuating short range order. (b) Out-of-plane momentum scans for (-3.0, 1.5, $\pm$0.5) along $L$. The scattering profile is fitted with two Lorentzian functions near $L=\pm0.5$ r.l.u. and a Gaussian function around $L=0$. (c) Hierarchy of correlation lengths. In-plane and out-of-plane correlation lengths for (-3.0, 1.5, 0.5) are shown. Here, $q^{\mathit{/}\!\mathit{/}}_\text{CDW}$ denotes the in-plane CDW wave vector. Solid lines represent fits to power-law dependence of the correlation length on $T - T_{\text{CDW}}$ above $T_{\text{CDW}}$. Below $T_{\text{CDW}}$, in the shaded gray region, the peak width corresponding to long-range order is limited by the instrument resolution. The dashed line represents an extension of the power-law fit to $T_{\text{CDW}}$, showing the diverging behavior. Below $T_{\text{CDW}}$, the correlation lengths are obtained from Gaussian fits to the peaks, while above $T_{\text{CDW}}$, they are extracted from Lorentzian fits.}
    \label{fig3}
\end{figure}

To visualize the out-of-plane diffuse scattering, we examine ($H$, $L$) momentum maps with $K = 0.5$ r.l.u. (reciprocal lattice units) as shown in Fig.~\ref{fig2}. The white dashed line drawn in Fig.~\ref{fig1}(a) schematically depicts this cut with respect to the in-plane ellipses. Below the CDW transition temperature ($T=$180~K), the observed peaks in the $K = 0.5$ r.l.u. plane arise due to the CDW superlattice (Fig.~\ref{fig2}(a)). They are resolution-limited both in the in-plane and out-of-plane directions, as can be seen more quantitatively in Fig.~\ref{fig3}. (Data well below $T_{\text{CDW}}$ are presented in the Supplemental Material). CDW peaks sit at half-integer $L$ values which indicates that the phase of the charge ordering pattern alternates between adjacent layers.

As the temperature is increased above $T_{\text{CDW}}$, only CDW fluctuations in the form of diffuse scattering is observed. Specifically, scattering intensity from CDW fluctuations form in a `raindrop'-like pattern and sharp CDW peaks are no longer present. (The diffuse scattering streaking diagonally across the images are due to thermal diffuse scattering from a transverse acoustic phonon that is not relevant to the present study.) For fixed $H$, intensities of the `raindrop' features are modulated with a period of four reciprocal lattice units along $L$. These intensity modulations, both in the diffuse scattering and in the CDW peaks below $T_{\text{CDW}}$, arise from the geometric structure factor associated with scattering from the Se atoms. Because the Se planes are positioned roughly 1/4 and 3/4 along the $c$ axis in the conventional unit cell, scattering intensities correspondingly vary along $L$, spanning a period of roughly four reciprocal lattice units. The data can be thought of as complete rods along $L$ that are interrupted by this structure factor effect. We can thus interpret the rod-like scattering along $L$ as indicating that the CDW fluctuations become uncorrelated in the out-of-plane direction for $T\gg T_{\text{CDW}}$.

To better quantify the CDW correlation lengths, we take $H$ and $L$ cuts at different temperatures. In Fig.~\ref{fig3}(a) we show an $H$-cut of a $q_3$-type CDW peak with Miller indices (-3.0, 1.5, 0.5). At each temperature a constant background, arising from thermal diffuse scattering from phonons not relevant to the CDW, is subtracted from these cuts. At 190\,K, the CDW peak is resolution limited. Above $T_{\text{CDW}}$, the diffuse scattering intensity decreases and the diffuse features broaden with increasing temperature. However, even up to our highest measured temperature, $T=400$\,K, diffuse scattering can still be resolved.

Out-of-plane momentum cuts of a pair of $q_3$-type CDW peaks (-3.0, 1.5, $\pm$0.5) are shown in Fig.~\ref{fig3}(b). The CDW peaks at 190\,K are resolution limited but have tails arising from diffuse scattering (Data well below $T_{\text{CDW}}$ showing negligible residual intensity between CDW peaks are presented in Supplemental Material). Above $T_{\text{CDW}}$, the diffuse peaks broaden with increasing temperature, with weight primarily concentrated into a central plateau that constitutes the `raindrop' centered at $L = 0$ (see Supplemental Material). This raindrop implies that the CDW loses out-of-plane phase coherence, even though the in-plane diffuse scattering indicates that the order within layers remains phase coherent over a considerable length. At 280\,K, for example, the diffuse scattering possesses a width of  $\delta H = 0.05$ r.l.u., but is broad in the $L$ direction with width $\delta L > 1$ r.l.u. (see Fig.~\ref{fig3}(a) and (b)).

\begin{figure}
    \centering
    \includegraphics[width=\columnwidth]{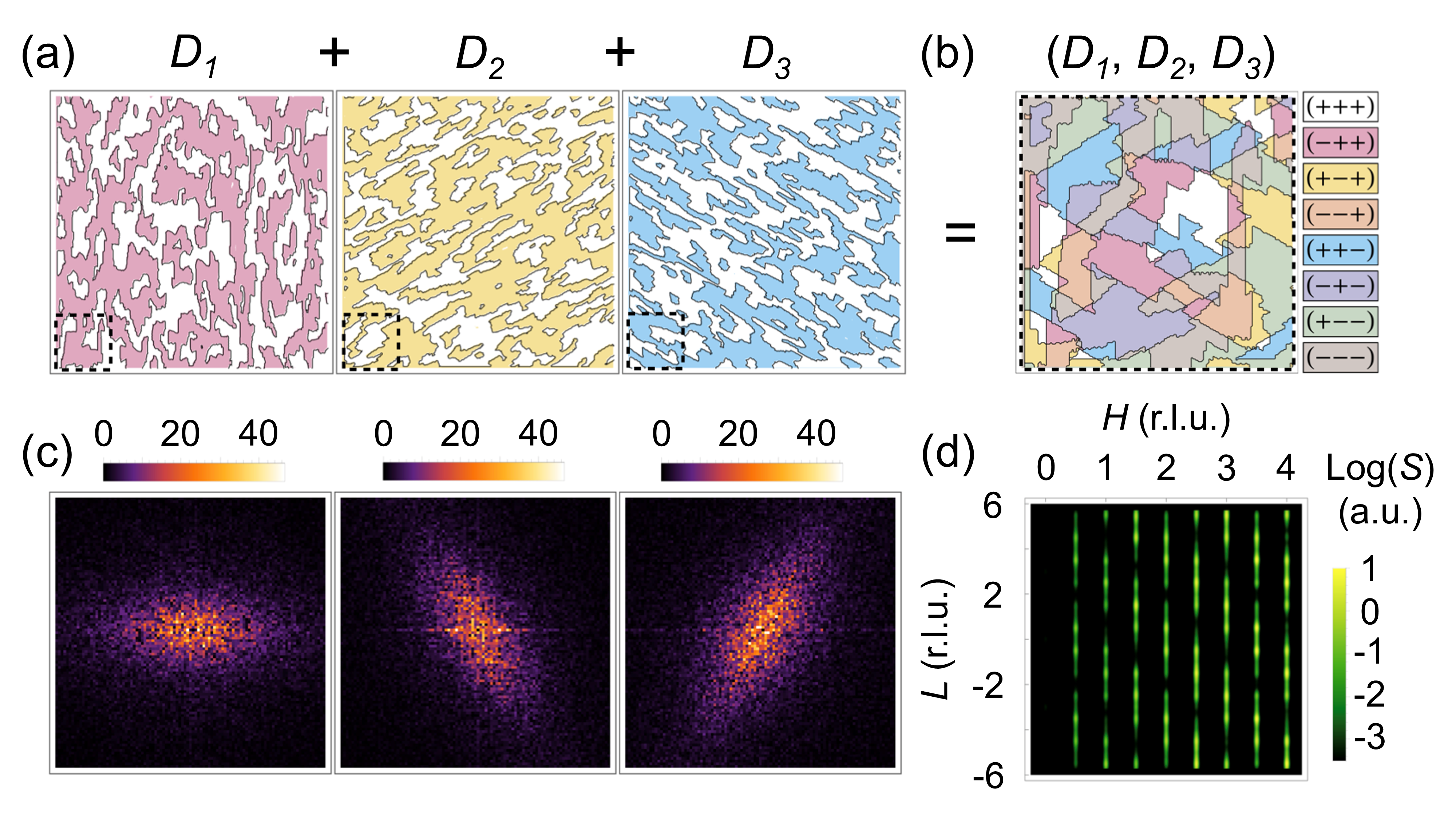}
    \caption{Structure factors from real space anisotropic CDW fluctuations. Because the charge order in 1$T$-TiSe$_2$ occurs in the strong coupling regime, the CDW amplitude is expected to be nonzero throughout the sample well above $T_{\text{CDW}}$. Finite correlation lengths in individual CDW components originate in domain walls across which individual CDW components invert (equivalent to a $\pi$ phase slip consistent with the 2$\times$2$\times$2 order). (a) In-plane domain distributions for the three CDW components obtained from modeling domains by an anisotropic Ising model (see Supplemental Material for details). (b) Overlapping the smaller dashed windows marked in the three plots in (a) yields the overall CDW distribution, with eight types of domains. (c) Simulated diffuse scattering from CDW fluctuations at each of the three in-plane CDW vectors $q_i$, obtained by Fourier transforming the corresponding plot in (a) and zooming in around the origin. The anisotropy in the scattering intensity reproduces the experimental observations shown in Fig.~\ref{fig1}(b). (d) Simulated $(H,L)$ structure factor for $K=0.5$ r.l.u., in the presence of a random stacking of alternating types of domains (see Supplemental Material for details). The intensity modulations of the  `rods' reproduce the observed modulations shown in Fig.~\ref{fig2}(c). 
    }
    \label{fig4}
\end{figure}

To quantify the anisotropy in the fluctuation regime, we summarize the hierarchy of correlation length scales in Fig.~\ref{fig3}(c). In this figure, we plot the in-plane and out-of-plane correlation lengths of a $q_3$-type CDW peak (-3.0, 1.5, 0.5) as a function of temperature above $T_{\text{CDW}}$. The fluctuations of each CDW component are clearly anisotropic in the plane, with the longer (shorter) correlation lengths in the direction perpendicular (parallel) to the in-plane wave vector of the corresponding CDW component. Meanwhile, the out-of-plane fluctuations have much shorter correlation lengths than in-plane ones above $T_{\text{CDW}}$. At 280~K, the CDW order in real space is everywhere locally triple-$q$ in nature, but each CDW component contains domain walls across which the CDW modulation pattern is inverted (i.e. a $\pi$ phase slip). The characteristic length scale between such domain walls is about 22 unit cells perpendicular to the in-plane wave vector of the CDW component, 7 unit cells parallel to the in-plane wave vector, and 2 unit cell in the out-of-plane direction (Fig.~\ref{fig3}(c)). This anisotropy is inherent within each individual CDW component, such that global three-fold rotational symmetry is preserved by the combination of three CDW components in the overall fluctuating triple-$q$ CDW order, as schematically depicted in Fig.~\ref{fig4}(a) and (b). Note that the correlation lengths along both in-plane and out-of-plane directions can be seen to all diverge at the same critical temperature $T_{\text{CDW}}$ by fitting their temperature dependence above $T_{\text{CDW}}$ to power laws in $T - T_{\text{CDW}}$, as shown in Fig.~\ref{fig3} (see Supplemental Material for details).

The simple picture that emerges from our studies is captured in Fig.~\ref{fig4}. Similar to most CDW-forming transition metal dichalcogenides, 1$T$-TiSe$_2$ lies in the strong coupling limit with 2$\Delta(0)$/k$_B T_{\text{CDW}} \approx 8.7$~\cite{Li2007}. It is expected, therefore, that a nonzero CDW amplitude is present throughout the sample in the temperature regime $T_{\text{CDW}} \lesssim T \ll T_\text{MF} \approx 500$\,K, without it being phase coherent over long distances~\cite{NbSe2_1,NbSe2_2}. The correlation lengths in each CDW component are determined by the mean distance between phase slips. These phase slips are dynamic so that neither short- nor long-range static order is present above $T_{\text{CDW}}$, except in regions near impurities, which can pin CDW order~\cite{Arguello2014, chatterjee2015emergence}. Because the phase slips occur independently for individual CDW components (see Supplemental Material), neighboring, fluctuating triple-$q$ regions in real space typically differ by a $\pi$ phase flip of just one component. Combining three CDW components, a total of eight different triple-$q$ fluctuation domain types can form. Many such domains together (Fig.~\ref{fig4}(b)) result in the anisotropic diffuse scattering intensities shown in Fig.~\ref{fig4}(c), which reproduces the experimental observation in Fig.~\ref{fig1}. Similarly, Fig.~\ref{fig4}(d) shows calculated structure factors in the $(H,L)$ plane in the presence of randomly stacked, but in-plane-coherent CDWs (see Supplemental Material for details), which reproduces the `raindrop' pattern of Fig.~\ref{fig2}.

Our results illustrate how a simple, single-component 2D-to-3D crossover picture is inadequate in describing the emergence of long-range order from local fluctuations in 1$T$-TiSe$_2$ and possibly in other CDW-forming transition metal dichalcogenides. The CDWs are instead characterized by anisotropic domains within CDW components. A hierarchy of length scales combines to yield a highly structured pattern of fluctuations upon approaching the transition temperature. Importantly, the fluctuations persist to high temperatures, and may therefore play an important role in the thermodynamic and electronic properties of these materials.

~\\
\emph{Acknowledgments}
\begin{acknowledgments}
We thank Doug Robinson for his help with our x-ray diffraction measurements. X-ray experiments were supported by the U.S. Department of Energy, Office of Basic Energy Sciences Grant No. DE-FG02-06ER46285 (P.~A.) and Grant No. DE-SC0023017 (A.~K.). F.~F.~acknowledges support from the Engineering and Physical Sciences Research Council, Grant No.~EP/X012239/1. F.~J.~acknowledges support from grant PID2021-128760NB0-I00 from the Spanish MCIN/AEI/10.13039/501100011033/FEDER, EU. P.~A.~acknowledges the Gordon and Betty Moore Foundation EPiQS Initiative through Grant No.~GBMF9452. Research at the University of Maryland was supported by the Gordon and Betty Moore Foundation's EPiQS Initiative through Grant No.~GBMF9071, the U.~S.~National Science Foundation (NSF) Grant No.~DMR2303090, and the Maryland Quantum Materials Center. T.~C.~C.~is supported by the U.~S.~Department of Energy (DOE), Office of Science (OS), Office of Basic Energy Sciences, Division of Materials Science and Engineering, under Grant No.~DE-FG02-07ER46383. G.~K.~acknowledges support by the National Science Foundation under Grant No.~ECCS-1711015. Work at Argonne (M.~J.~K., S.~R., R.~O.: data collection and reduction) was supported by the U.~S.~Department of Energy, Office of Science, Basic Energy Sciences, Materials Sciences and Engineering Division. Use of the Advanced Photon Source at Argonne National Laboratory was supported by the US Department of Energy, Office of Science, Office of Basic Energy Sciences under contract No.~DE-AC02-06CH11357.
\end{acknowledgments}

\bibliography{mylib}

\end{document}


\title{Supplemental Material for \\ In-plane anisotropy of charge density wave fluctuations in 1$T$-TiSe$_2$}

\author{Xuefei Guo}
\email{xuefeig2@illinois.edu}
\affiliation{\UIUC}
\author{Anshul Kogar}
\email{anshulkogar@physics.ucla.edu}
\affiliation{\UCLA}
\author{Jans Henke}
\affiliation{\Amsterdam}
\author {Felix Flicker}
\affiliation{\Bristol}
\author {Fernando de Juan}
\affiliation{\Donostia}
\affiliation{\IKERBASQUE}
\author{Stella X.-L. Sun}
\affiliation{\UIUC}
\author{Issam Khayr}
\affiliation{\UIUC}
\affiliation{\UMN}
\author{Yingying Peng}
\affiliation{\PKU}
\author{Sangjun Lee}
\affiliation{\UIUC}
\author{Matthew J. Krogstad}
\author{Stephan Rosenkranz}
\author{Raymond Osborn}
\affiliation{\APS}
\author{Jacob P. C. Ruff}
\affiliation{\CHESS}
\author{David B. Lioi}
\author{Goran Karapetrov}
\affiliation{\DU}
\author{Daniel J. Campbell}
\affiliation{\UMD}
\author{Johnpierre Paglione}
\affiliation{\UMD}
\affiliation{\CIFAR}
\author{Jasper van Wezel}
\affiliation{\Amsterdam}
\author{Tai C. Chiang}
\author{Peter Abbamonte}
\email{abbamont@illinois.edu}
\affiliation{\UIUC}

\date{\dateOfSubmission}


\maketitle
\section{Determining $T_{\text{CDW}}$ of the semimetallic sample}
\begin{figure}
    \centering
    \includegraphics[scale=0.5]{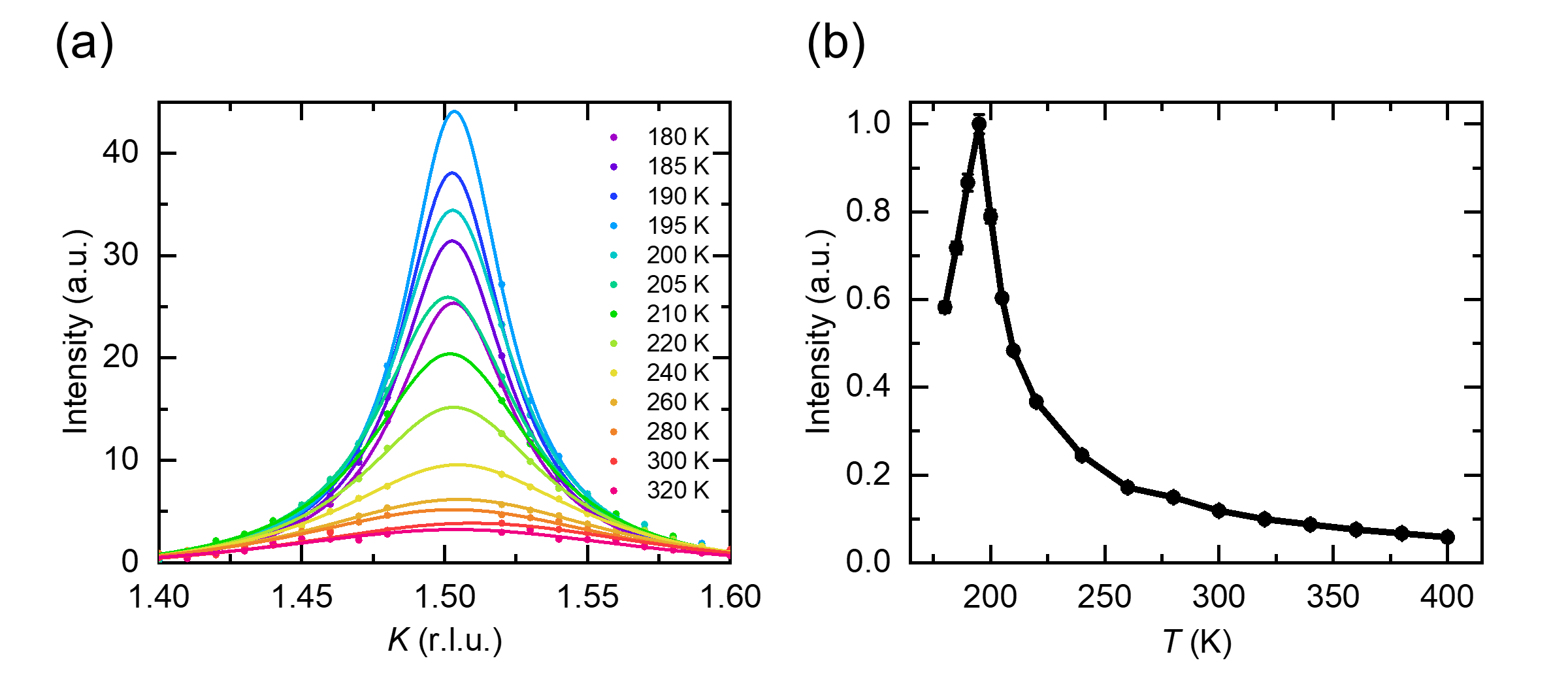}
    \caption{Thermal diffuse scattering of the CDW peak at (-3.0, 1.5, 0.5). (a) Line cuts along $K$ with the Bragg peak subtracted. The thermal diffuse scattering increases as the temperature rises to $T_{\text{CDW}}$ and then decreases as the sample is further heated. This trend is evident from the raw data points without requiring a fit. The lines represent Lorentzian fits. (b) The peak intensity of the Lorentzian functions shown in (a), with the intensity reaching a maximum at $T_{\text{CDW}}=195$\,K.}
    \label{Tc_diffuse}
\end{figure}

Thermal diffuse scattering of the charge density wave (CDW) peak with Miller indices (-3.0, 1.5, 0.5) is modeled using Lorentzian functions, both below and above the CDW transition temperature $T_{\text{CDW}}$. The Bragg peak region is excluded, which corresponds to 1.49 - 1.51 r.l.u. in $K$ as shown in Fig.~\ref{Tc_diffuse}(a). The diffuse scattering intensity shows a peak at a particular temperature, which is defined as $T_{\text{CDW}}=195$\,K as illustrated in Fig.~\ref{Tc_diffuse}(b).

\section{High-energy x-ray diffuse scattering analysis of the semimetallic sample}
\begin{figure}[bth]
    \centering
    \includegraphics[scale=0.5]{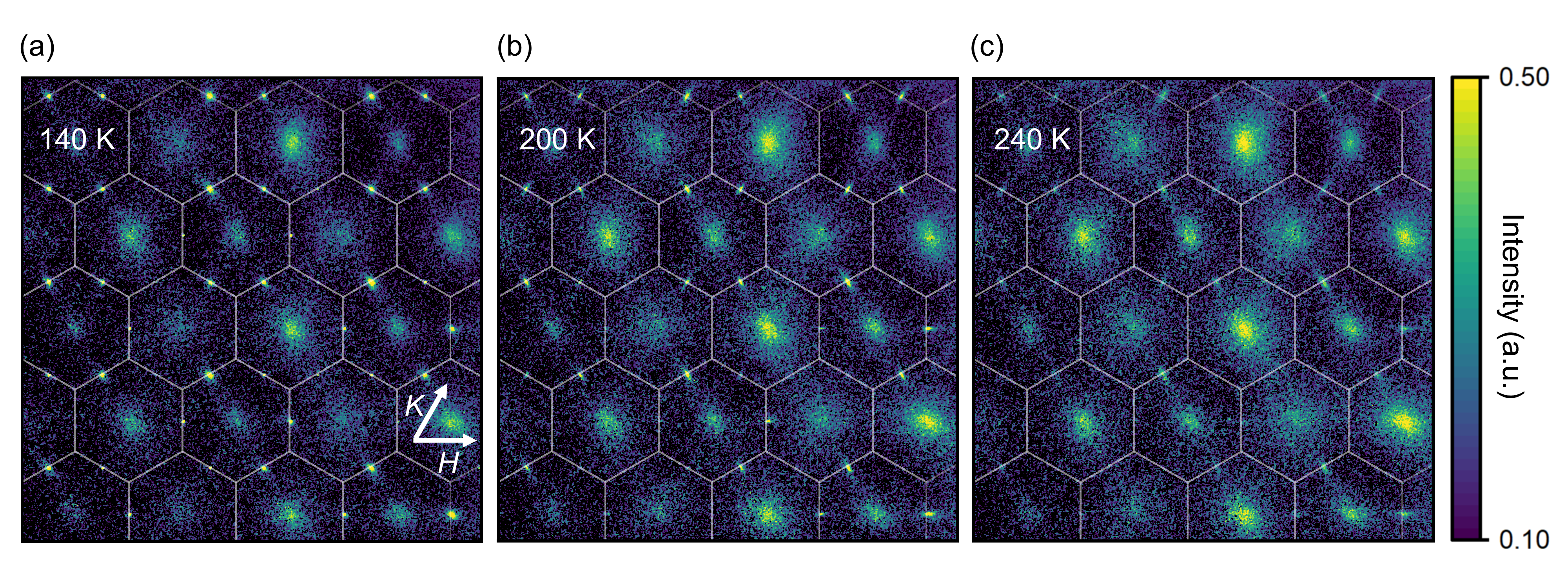}
    \caption{In-plane momentum map of the CDW at $L=0.5$ r.l.u. at (a) 140\,K (well below $T_{\text{CDW}}$), (b)200\,K (just above $T_{\text{CDW}}$) and (c) 240\,K (above $T_{\text{CDW}}$). At the base temperature, the CDW peaks are resolution-limited. Above $T_{\text{CDW}}$, the CDW fluctuations exhibit in-plane anisotropy, with the peak elongated along the $q^{\mathit{/}\!\mathit{/}}_\text{CDW}$ direction. The hexagonal white lines represent the normal state Brillouin zone. The scale bar is in arbitrary units and follows a logarithmic scale.}
    \label{in-plane_map}
\end{figure}

High energy x-rays with an energy of 56.7\,keV are used to probe a larger momentum range of the CDW diffuse scattering on the semimetallic sample. Sharp, resolution-limited CDW peaks are observed at 140\,K, well below $T_{\text{CDW}}$. At 200\,K, just above $T_{\text{CDW}}$, diffuse tails parallel to the in-plane $q_{\text{CDW}}$ are identified, as shown in Fig.~\ref{in-plane_map}(b). At 240\,K, well above $T_{\text{CDW}}$, only diffuse scattering remains at the $q_{\text{CDW}}$ positions, with a clear in-plane anisotropy. The in-plane momentum map covering a broader momentum range as shown in Fig.~\ref{in-plane_map}, is consistent with the one presented in Fig. 1 of the main manuscript.

\begin{figure}
    \centering
    \includegraphics[scale=0.5]{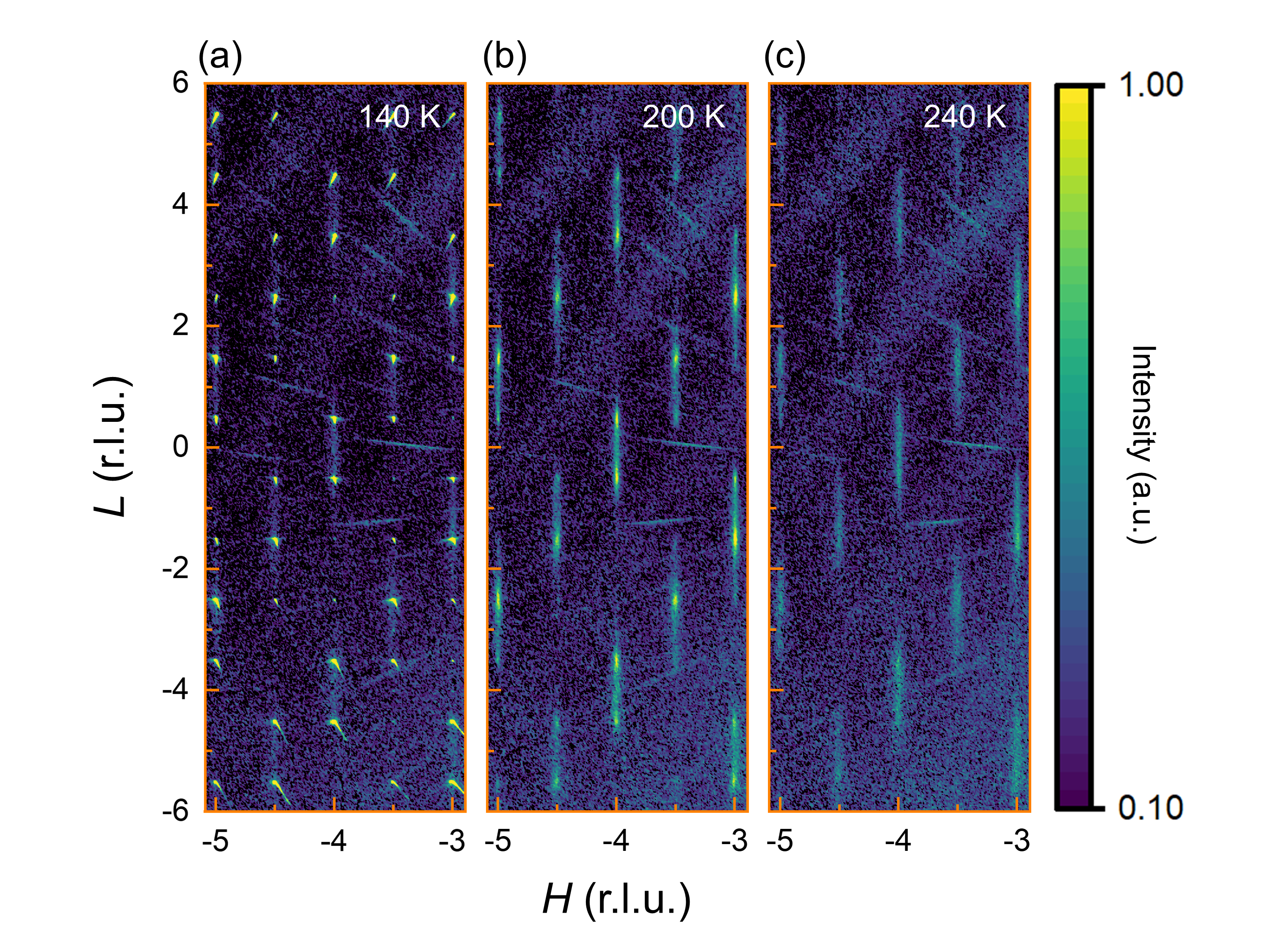}
    \caption{Out-of-plane momentum map of the CDW at $K=0.5$ r.l.u. at (a) 140\,K (well below $T_{\text{CDW}}$), (b)200\,K (just above $T_{\text{CDW}}$) and (c) 240\,K (above $T_{\text{CDW}}$). Below $T_{\text{CDW}}$, the CDW peaks are resolution-limited with the tails arising from the mosaicity of the sample. The ``raindrop" pattern observed above $T_{\text{CDW}}$ is the same as that shown in Fig. 2 of the main manuscript.}
    \label{out-of-plane_map}
\end{figure}

In the out-of-plane momentum map at $K=0.5$ r.l.u. shown in Fig.~\ref{out-of-plane_map}(a), all sharp features correspond to the CDW peaks. The sharp streaks seen in the figures are the artifacts from the detector due to the saturation of peaks from neighboring planes. At 200\,K, slightly above $T_{\text{CDW}}$, raindrop-like scattering is the dominant feature, although peak residual remains at the $q_{\text{CDW}}$ positions. At 240\,K, well above $T_{\text{CDW}}$, only fluctuation patterns remain, as shown in Fig.~\ref{out-of-plane_map}(c). These patterns are again consistent with those presented in Fig. 2 of the main manuscript. 

\begin{figure}
    \centering
    \includegraphics[scale=0.5]{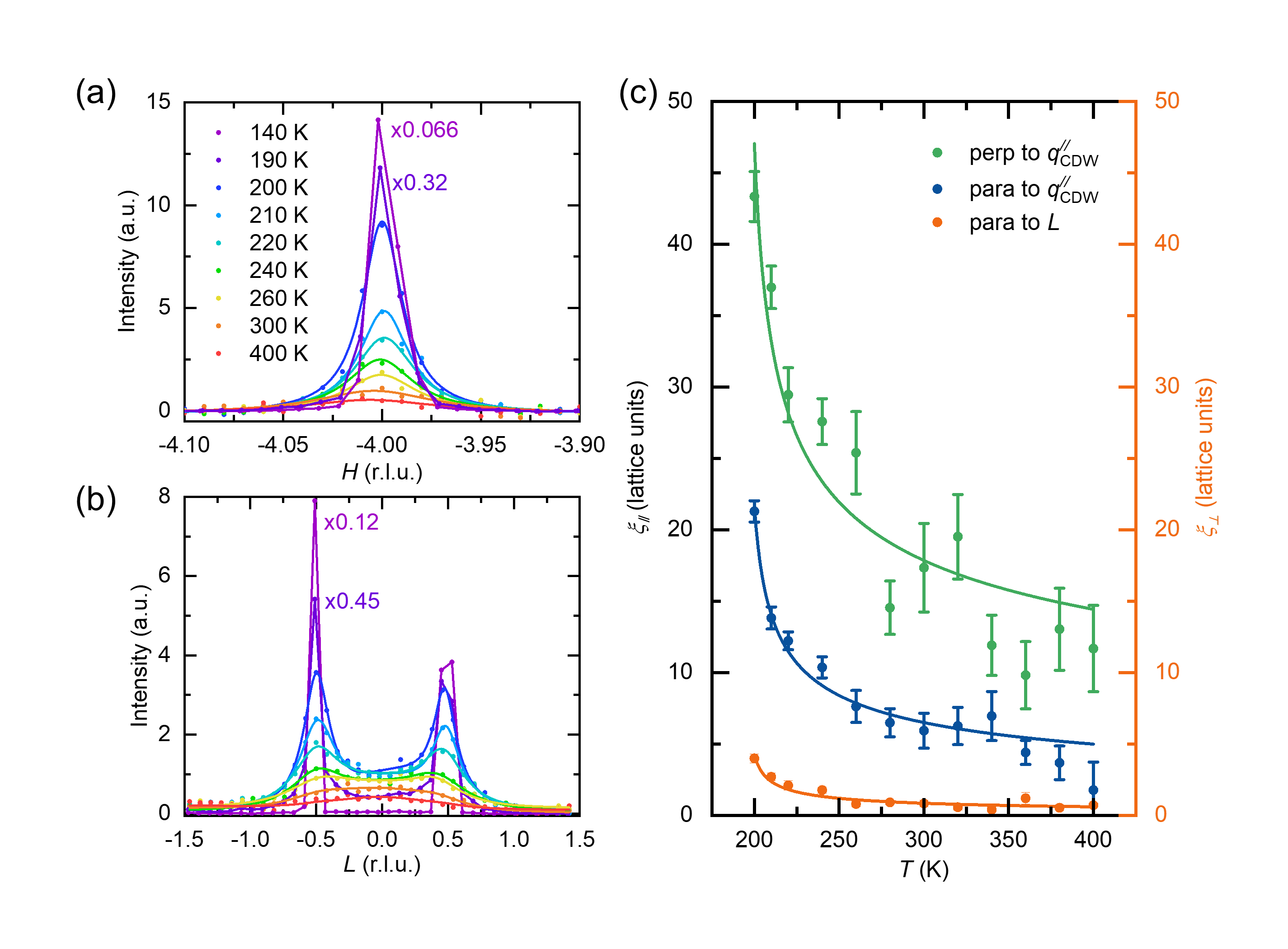}
    \caption{Line cuts of the CDW peaks and their correlation lengths. (a) Momentum scans of the CDW peak at (-4.0, 1.5, -2.5) along the $H$-direction, both below and above $T_{\text{CDW}}$. The CDW peak is sharp at the base temperature, while the diffuse scattering is broad above $T_{\text{CDW}}$. (b)$L$-cuts of the pair of CDW peaks at (-3.0, 1.5, $\pm$0.5). At the base temperature, there is negligible weight between the half-integer CDW peaks. As the temperature increases, spectral weight begins to accumulate at $L=0$, and the overall thermal diffuse scattering weakens. (c) In-plane and out-of-plane correlation lengths of CDW diffuse scattering at (-4.0, 1.5, -2.5). The lines are power-law fits of $T-T_{\text{CDW}}$.}
    \label{line_cut}
\end{figure}

Detailed line scans of the CDW fluctuations are presented in Fig.~\ref{line_cut}. At the base temperature of 140\,K, a sharp CDW peak at (-4.0, 1.5, -2.5) along $H$ is shown in Fig.~\ref{line_cut}(a). At 190\,K, just below $T_{\text{CDW}}$, resolution-limited peak is still present, but a thermal diffuse scattering tail becomes visible in the data. At 200\,K, just above $T_{\text{CDW}}$, the sharp peak now is not prominent, and the thermal diffuse scattering can be fitted with a Lorentzian function. As the temperature increases further, the intensity of the thermal diffuse scattering progressively weakens. 

In the out-of-plane direction at (-3.0, 1.5, $\pm$0.5), as shown in Fig.~\ref{line_cut}(b), sharp peaks are observed at the half-integer positions at the base temperature, with little to no visible tails between the peaks. At 190\,K, in addition to the CDW peaks, tails begin to emerge, and fluctuations appear on both sides of the CDW peaks. Just above $T_{\text{CDW}}$, at 200\,K, substantial weight is observed around $L=0$ r.l.u.. As the temperature increases further, the peaks at the half-integer positions begin to diminish, and the feature evolves into a raindrop-like pattern.

The hierarchy of the correlation lengths for (-4.0, 1.5, -2.5) is summarized in Fig.~\ref{line_cut}(c), where the correlation length perpendicular to the in-plane $q_{\text{CDW}}$ is larger than that parallel to it. Additionally, the out-of-plane correlation length is significantly smaller than the in-plane one. All three temperature-dependent correlation lengths are well described by a power-law function of $T-T_{\text{CDW}}$, with $T_{\text{CDW}}=195$\,K.

\newpage
\section{Hard x-ray diffraction analysis of the vacancy-reduced sample}
A similar experiment is done on the vacancy-reduced sample with a photon energy of 87.4\,keV at the Advanced Photon Source Sector 6-ID-D.

\begin{figure}
    \centering
    \includegraphics[scale=0.5]{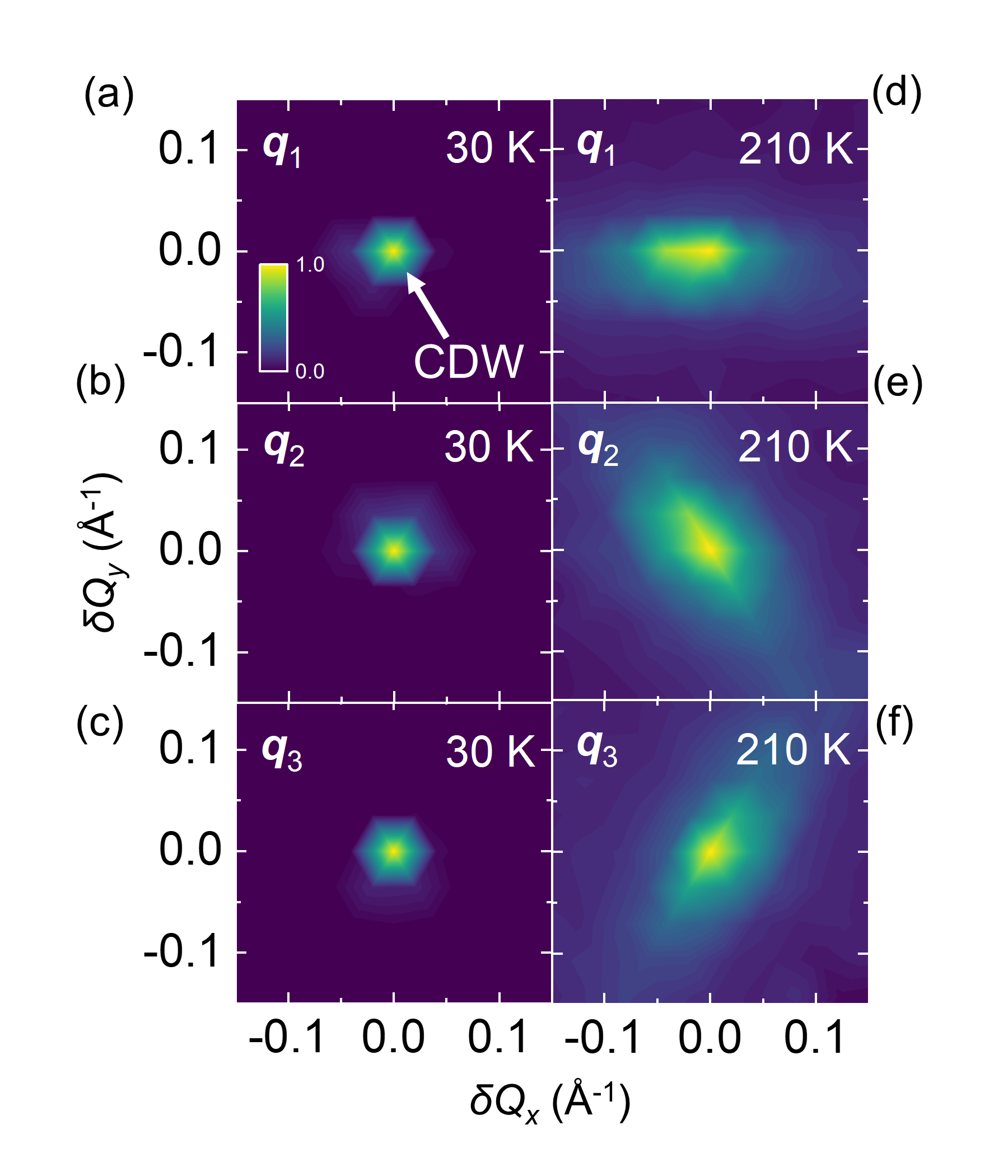}
    \caption{In-plane momentum maps around the CDW regions of the vacancy-reduced sample. (a)-(c) Scattering peaks associated with CDW components of $q_1$ type (2.5, -6.0, -1.5), $q_2$ type (0.5, -6.5, -1.5) and $q_3$ type (-7.0, 0.5, -0.5) at 30\,K. $q_i$ is defined in Fig. 1 of the main manuscript and $Q_x$ is parallel to $H$. $\delta Q_x$ and $\delta Q_y$ are the distances away from the CDW peak position. The CDW peaks are resolution limited. (d)-(f) The same momentum regions as (a)-(c) respectively, but at 210\,K, which lies above $T_{\text{CDW}}$=208\,K. Intensity scales have been normalized to the maximum intensity in the respective color plots.}
    \label{in-plane_APS}
\end{figure}

In Figs.~\ref{in-plane_APS}(a)-(c) we show that the CDW peaks are resolution-limited at the base temperature of 30\,K. The three peaks represent the three distinct CDW components; their wave vectors are of $q_1$-type (2.5, -6.0, -1.5) (Fig.~\ref{in-plane_APS}(a)), $q_2$-type (0.5, -6.5, -1.5) (Fig.~\ref{in-plane_APS}(b)) and $q_3$-type (-7.0, 0.5, -0.5) (Fig.~\ref{in-plane_APS}(c)), where ($H$, $K$, $L$) represent Miller indices. As the sample temperature is increased to 210\,K, just above $T_{\text{CDW}}$, CDW fluctuations become clearly visible. The anisotropic pattern observed is the same as that seen in the semimetallic samples, as shown in Fig.~\ref{in-plane_map}.

\begin{figure}
    \centering
    \includegraphics[scale=0.5]{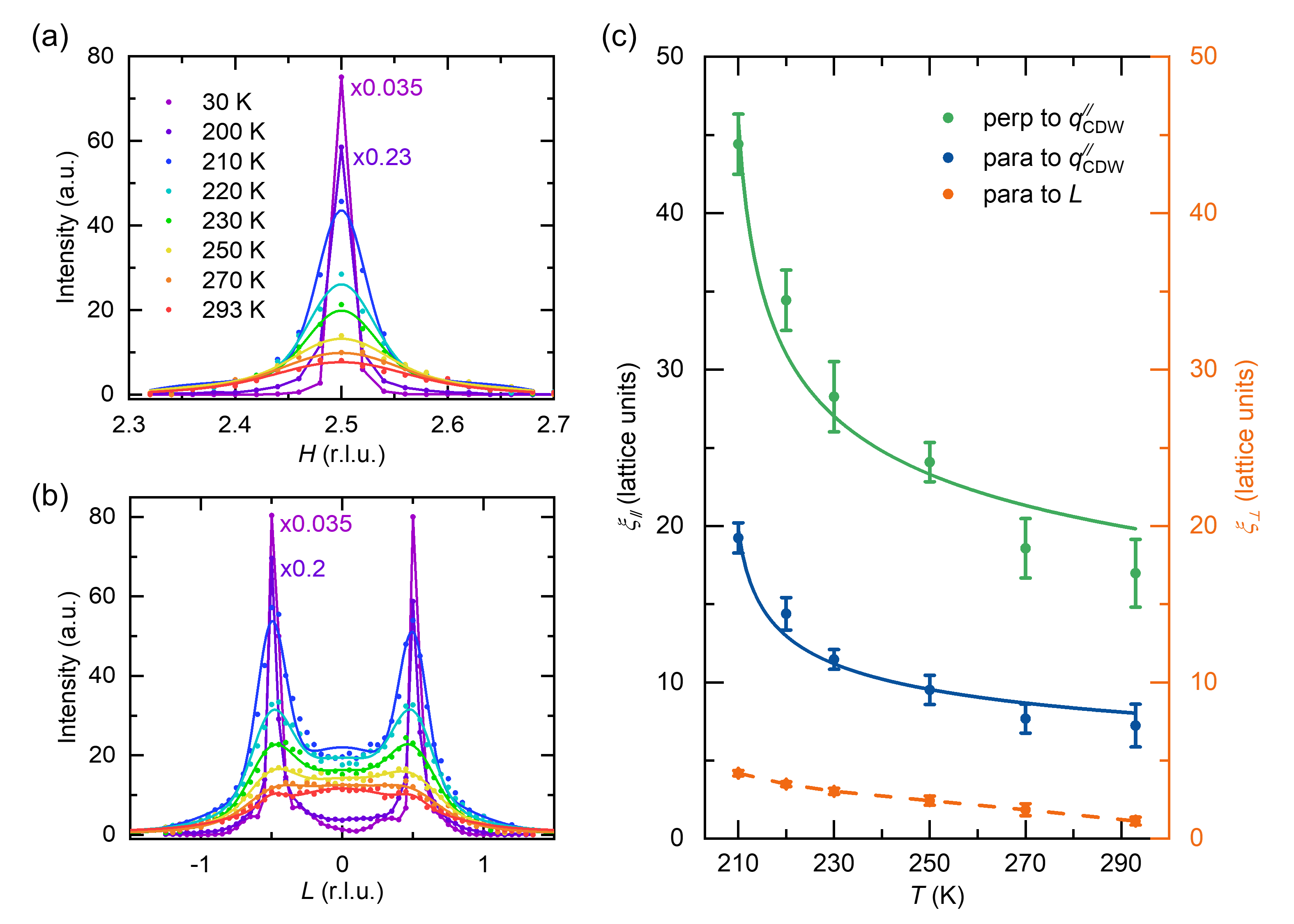}
    \caption{Momentum scans and correlation lengths of the vacancy-reduced sample. (a) $H$-cuts of the CDW peaks at (2.5, -5.0, 0.5) and this cut is parallel to the in-plane CDW wave vector. Strong diffuse scattering remains present at 293\,K for the vacancy-reduced sample. Above $T_{\text{CDW}}$, the peak is fitted with a power-law function convolved with an instrumental Gaussian function. (b) $L$-cuts of the CDW pair at (4.0, -3.5, $\pm$0.5). Above $T_{\text{CDW}}$, the spectral weight begins to shift from half-integer values to the center at $L=0$ r.l.u.. The solid line is a fit with three Gaussian functions. (c). The hierarchy of the correlation lengths for the CDW fluctuations at (5.0, -5.5, -0.5). The solid lines are power-law fits of the $T-T_{\text{CDW}}$ while the dashed line serves as a guide to the eye.}
    \label{line_cut_APS}
\end{figure}

In Fig.~\ref{line_cut_APS}(a), an $H$-cut of a $q_1$-type CDW peak at (2.5, -5.0, 0.5) is shown; this cut is parallel to the in-plane CDW wave vector. At the base temperature of 30\,K, the CDW peak is resolution-limited. At 200\,K, just below the $T_{\text{CDW}}$, a tail near the CDW peak becomes apparent in the data. As the temperature increases, the intensity of the diffuse scattering gradually decreases. The overall feature is consistent with the observations made for the semimetallic sample as shown in Fig.~\ref{line_cut}(a).

\begin{figure}
    \centering
    \includegraphics[scale=0.5]{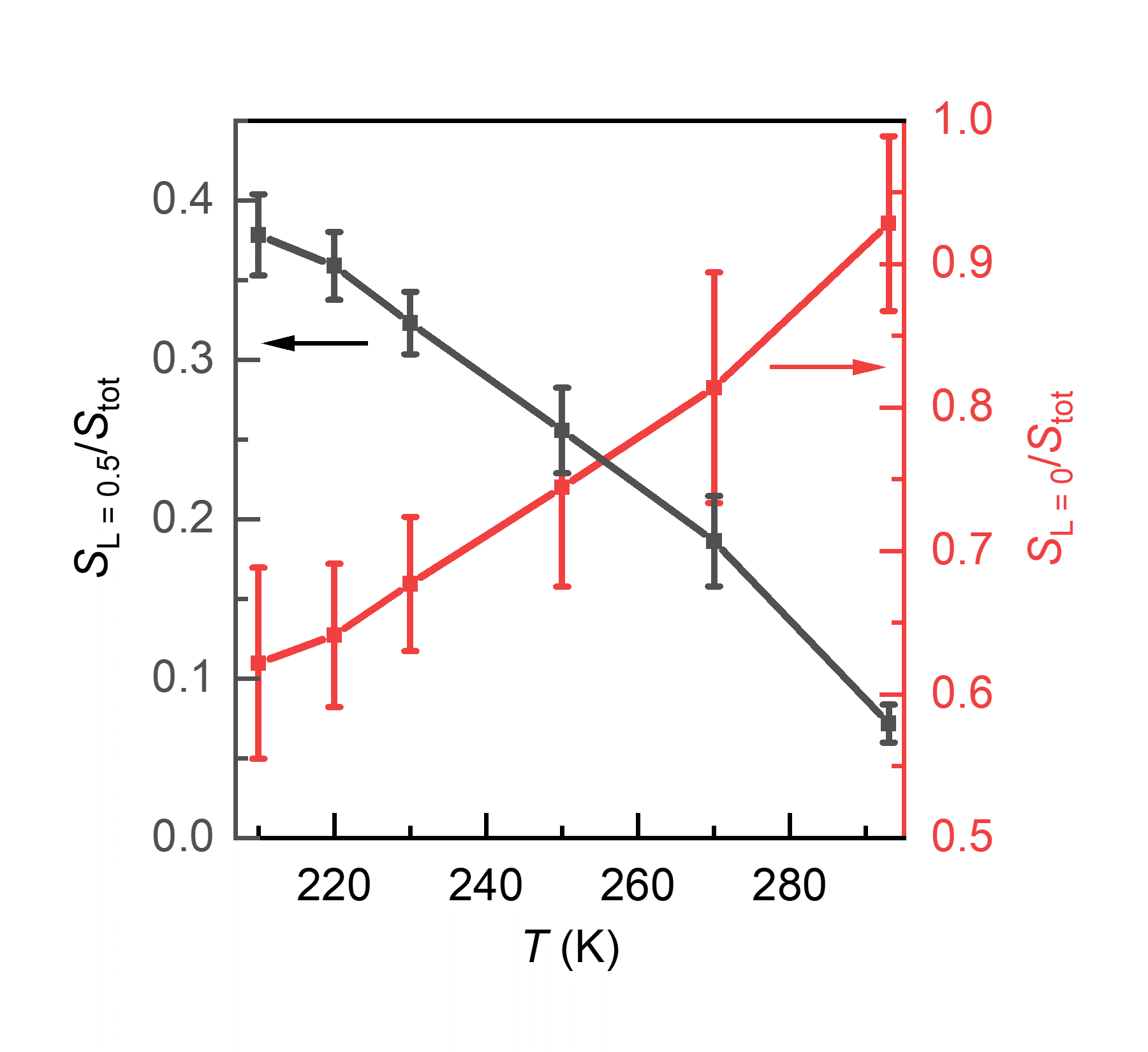}
    \caption{The weight transfer from half-integer to integer values, for the peaks at (4.0, -3.5, $\pm$0.5) along $L$. The three peaks centered at $L=\pm0.5$ r.l.u. and $L=0$ are fitted with three Gaussian functions. The peak area under $L=0.5$ r.l.u. ($S_{\mathrm{L=0.5}}$), $L=0$ ($S_{\mathrm{L=0}}$), and the total area $S_{\mathrm{tot}}$ are then evaluated from these Gaussian fits.}
    \label{spectral_transfer}
\end{figure}

Fig.~\ref{line_cut_APS}(b) shows $L$ cuts of $q_3$-type CDW peaks at (4.0, -3.5, $\pm$0.5). At the base temperature, the resolution-limited CDW peaks exhibit tails arising from stacking faults. At 200\,K, just below $T_{\text{CDW}}$, a noticeable spectral weight around $L=0$ r.l.u. is observed. As the temperature further increases above the $T_{\text{CDW}}$, CDW is melted, but strong diffuse scattering remains in Fig.~\ref{line_cut_APS}(b) as a residual of the peaks. This is associated with a ``raindrop" pattern in the $H$-$L$ plane, where the peak remains relatively ``sharp" in $H$ while becoming broader in $L$. The sharp peaks observed at half-integer $L$ positions below $T_{\text{CDW}}$ begin to blur, and peak centroids shift from half-integer to integer values, indicating that the out-of-plane staggered structure of CDW order is disrupted, although the layers themselves remain fairly well-ordered. To better quantify the weight transfer, the area under the three Gaussian functions centered at $L=\pm 0.5$ r.l.u. and $L = 0$ is calculated. As temperature increases, weight starts to shift from $L=\pm0.5$ r.l.u. to $L=0$, as illustrated in Fig.~\ref{spectral_transfer}. 

The correlation lengths of (5.0, -5.5, -0.5) are presented in Fig.~\ref{line_cut_APS}. Noticeably, the values of the correlation lengths closely match those in the semimetallic samples, as shown in Fig.~\ref{line_cut}. This demonstrates that the anisotropic properties of the CDW fluctuations are not linked to the transport properties of the 1$T$-TiSe$_2$ sample, nor to the CDW peaks at specific momentum positions.

Near a phase transition, relevant physical quantities exhibit universal scaling behavior that is independent on microscopic details and is only dependent on the interaction range and system dimensionality. Below $T_{\text{CDW}}$, the integrated intensity of the CDW at $\mathbf{q}$, which is a measure of the square of the total electron density $\rho_{\mathbf{q}}$, is proportional to the square of the order parameter $\Delta$. This relationship can be used to determine the critical exponent $\beta$ and transition temperature $T_{\text{CDW}}$ near the phase transition. Above $T_{\text{CDW}}$, diffuse scattering can probe the correlation length $\xi$ of the fluctuations, inferred from the width of the diffuse peaks, thereby determining the critical exponent $\nu$. Additionally, the shape of the peak near the CDW wave vector $\mathbf{q}$, $I(\mathbf{q})$, represents the correlation function and can be used to measure the critical exponent $\eta$ near the phase transition. In summary, the critical exponents of interest in this study are
\begin{subequations}
\label{critical_exponent}
\begin{equation}
I(T) = I_0\left(1-\frac{T}{T_{\text{CDW}}}\right)^{2\beta} \label{Tc_eq}
\end{equation}
\begin{equation}
I(\mathbf{q})=I_{\mathbf{q}}\left(\mathbf{q}-\mathbf{q_c}\right)^{-2+\eta} \label{eta_eq}
\end{equation}
\begin{equation}
\xi(T)=\xi_0\left(\frac{T}{T_{\text{CDW}}}-1\right)^{-\nu} \label{xi_eq}
\end{equation}
\end{subequations}

\begin{figure}
    \centering
    \includegraphics[scale=0.5]{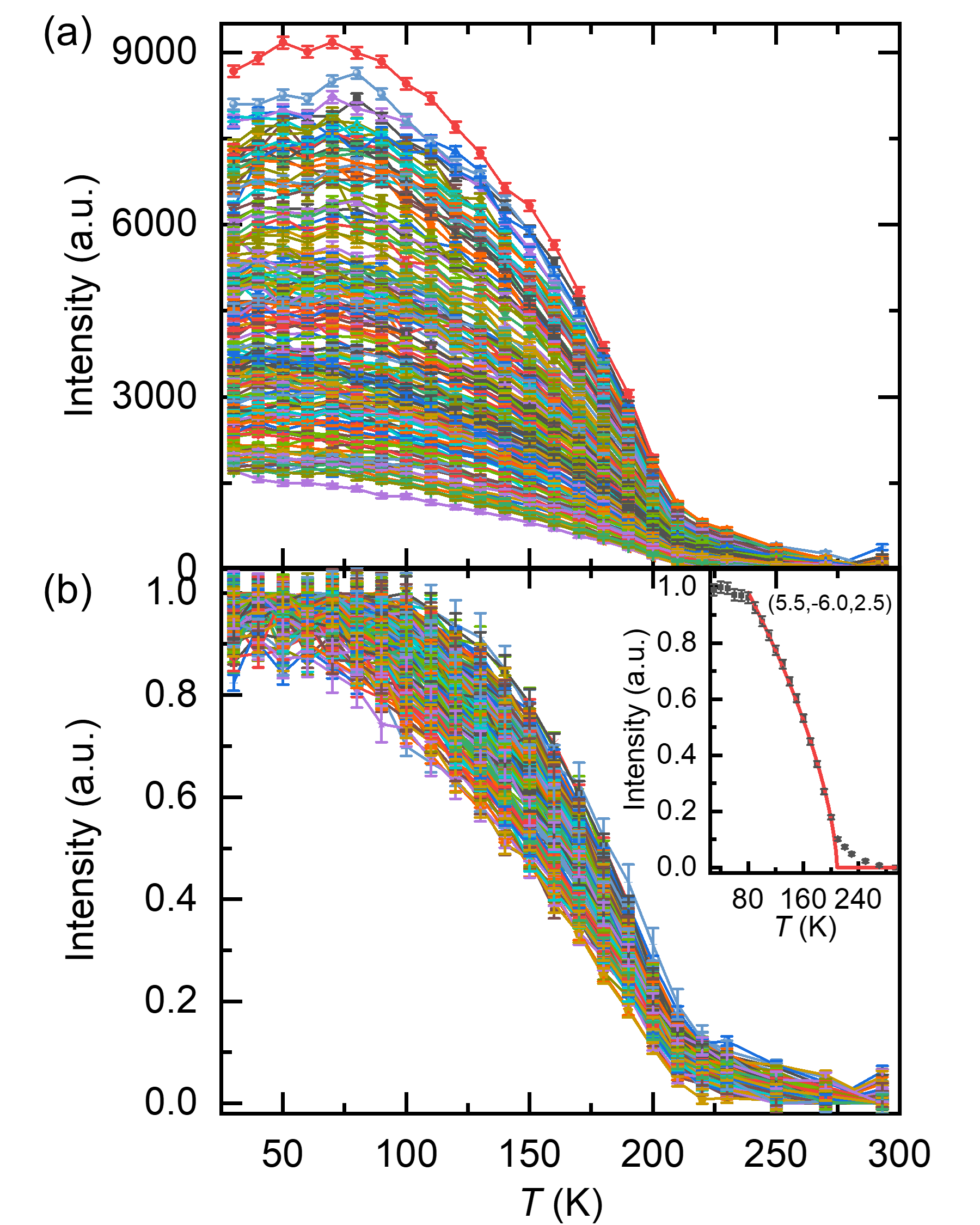}
    \caption{Integrated intensity of CDW peaks across the phase transition. (a) A broad range of CDW peak intensities is measured. (b) The scaled intensity at low temperature reveals a continuum of functional forms for $I(T)$. The inset shows a typical fitting curve for the CDW at (5.5, -6.0, 2.5). The pronounced tail near $T_{\text{CDW}}$ arises from a combination of crystallographic imperfections in the sample and the short-range fluctuations.}
    \label{Tc_fit}
\end{figure}

\begin{figure}
    \centering
    \includegraphics[scale=0.5]{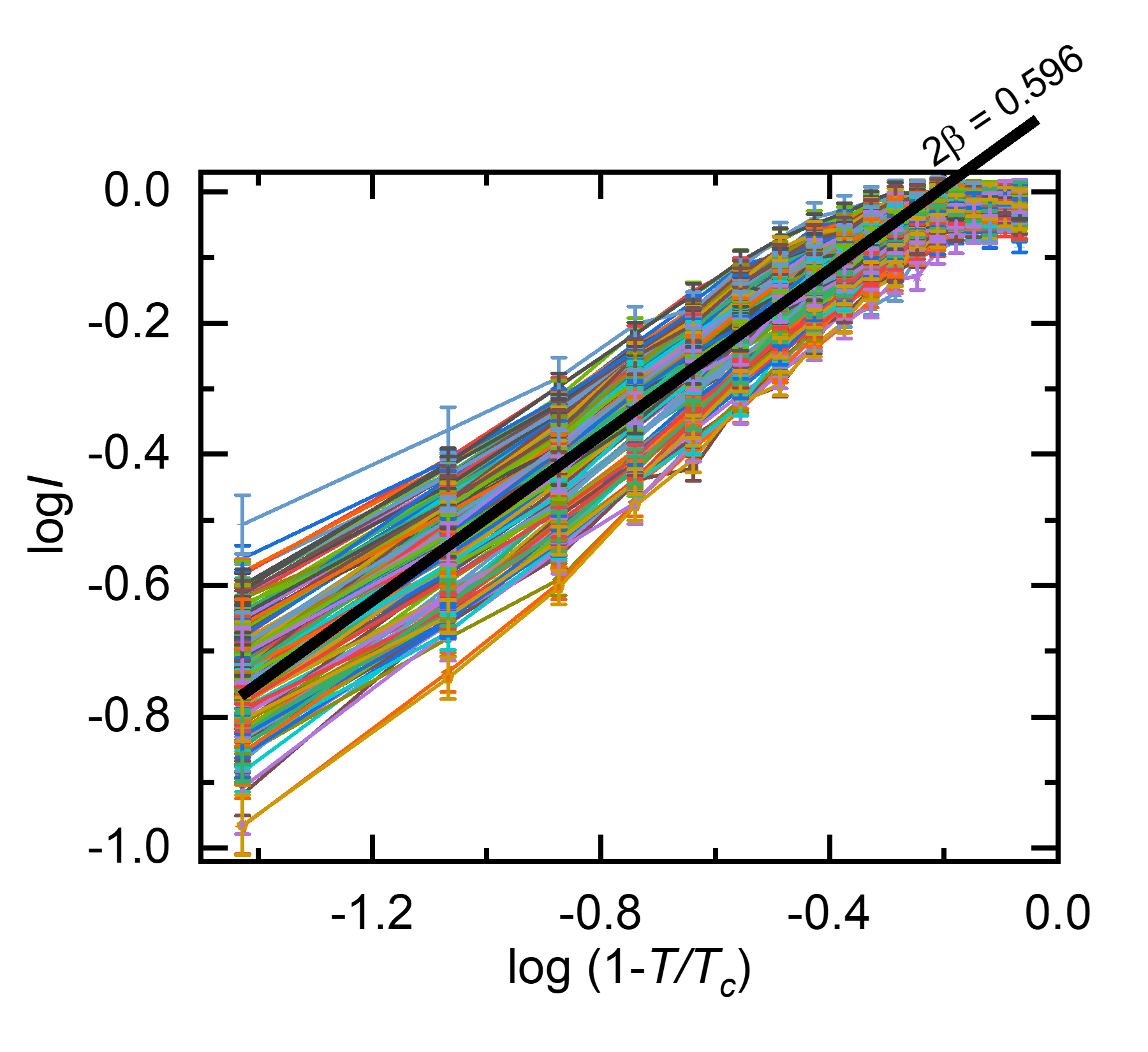}
    \caption{Log-log plot of CDW intensities versus temperature, illustrating the critical behavior near the phase transition. Near $T_{\text{CDW}}$, a power-law trend is observed before deviating at low temperatures.}
    \label{log-log_TC}
\end{figure}

\begin{figure}
    \centering
    \includegraphics[scale=0.4]{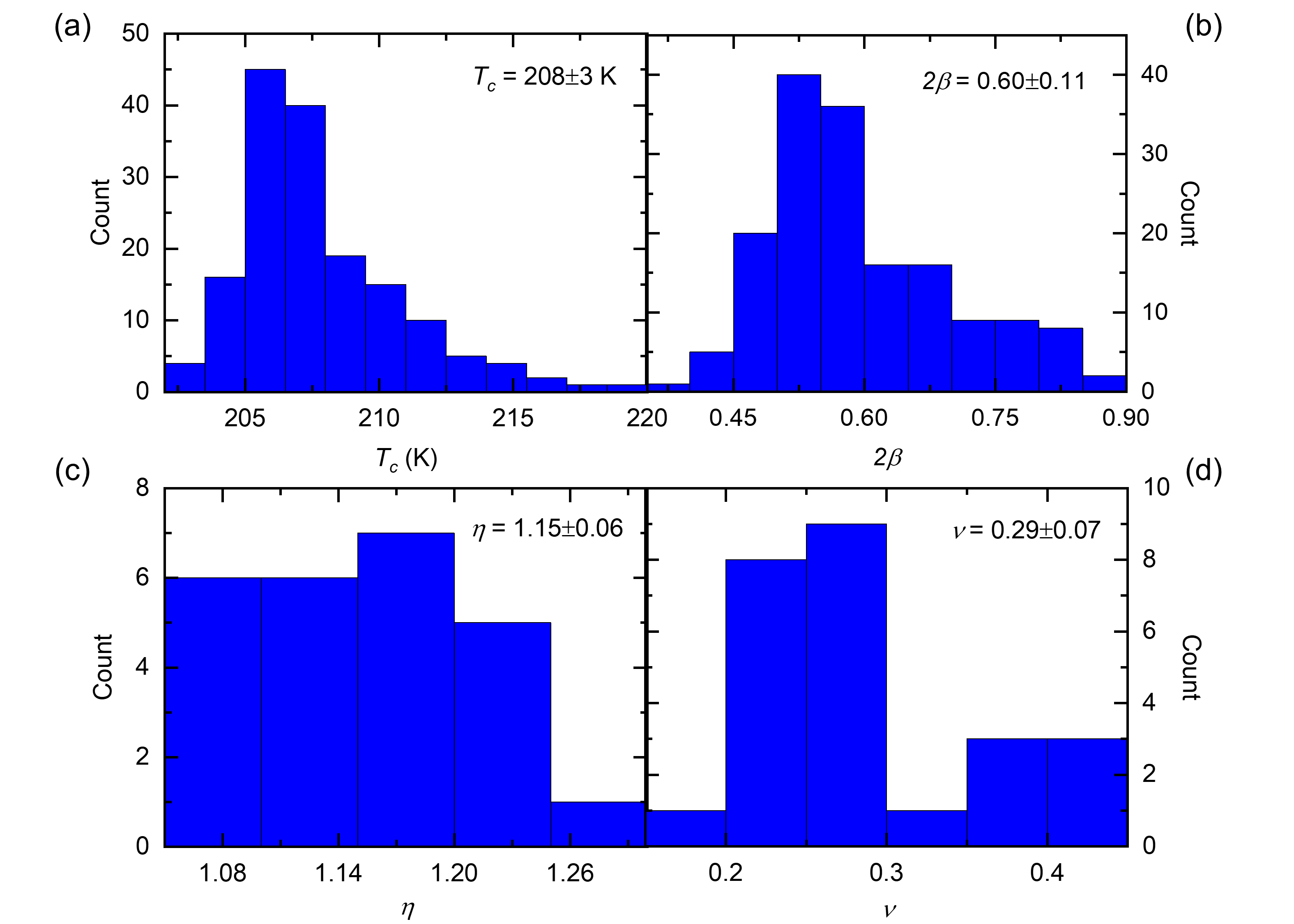}
    \caption{Histograms of $T_{\text{CDW}}$ and critical exponents. (a) CDW transition temperature $T_{\text{CDW}}$. (b)-(d) Critical exponents $\beta$, $\eta$ and $\nu$.}
    \label{hist}
\end{figure}

To determine $T_{\text{CDW}}$ and $\beta$ of the vacancy-reduced sample, we monitored the intensity $I(T)$ of hundreds of CDW peaks across the phase transition. A wide range of intensities was observed as shown in Fig.~\ref{Tc_fit}(a). To facilitate comparison with the functional form for $\beta$ in Eq.~\eqref{Tc_eq}, we scaled the intensity to its low-temperature value, allowing us to observe a continuum of functional forms near $T_{\text{CDW}}$ in Fig.~\ref{Tc_fit}(b). This variation in $\beta$ is more clearly illustrated in the log-log plot of intensity versus temperature in Fig.~\ref{log-log_TC}. A histogram summarizing the values of $T_{\text{CDW}}$ and $\beta$ is presented in Fig.~\ref{hist}(a) and (b), with $T_{\text{CDW}} = 208\pm3$ K and $2\beta = 0.60\pm0.11$.

The critical exponents $\eta$ and $\nu$ are determined from the shape of the CDW fluctuation peaks above $T_{\text{CDW}}$, as shown in Fig.~\ref{line_cut_APS}(a). To fit the in-plane CDW fluctuations, power-law functions described by Eq.~\eqref{eta_eq} are convolved with an instrumental Gaussian function, allowing for the extraction of $\eta$. The in-plane correlation lengths at different temperatures are shown in Fig.~\ref{line_cut_APS}(c), from which $\nu$ is calculated based on the in-plane correlation length. Only peaks unaffected by extraneous background scattering, strong mosaic rings and detector artifacts are used in these fittings. The histogram summarizing the in-plane CDW diffuse scattering of $\eta$ at $T = 210$ K and $\nu$ is presented in Fig.~\ref{hist}(c) and (d), with $\eta = 1.15\pm0.06$ and $\nu=0.29\pm0.07$.

The hyperscaling relation states that
\begin{equation}
d=\frac{2\beta}{\nu}+2-\eta
\end{equation}
where $d$ is the spatial dimension of the system. Based on the critical exponents estimated from our experiment, the effective dimension $d=2.9 \pm 0.6$, which aligns with the expectations for a quasi-2D system.

\section{Ginzburg-Landau theory of CDW order melting}
We present a Ginzburg-Landau theory describing the melting of CDW order in $1T$-TiSe$_2$ through the proliferation of phase fluctuations. The analysis predicts that at high temperatures, local anisotropic domains of triple-$Q$ CDW order are separated by domain walls across which one of the CDW components undergoes a $\pi$-phase shift.

\subsection{The ground state}\label{sec:gs}

The CDW phase in $1T$-TiSe$_2$ consists of the modulation $\alpha(\mathbf{r})$ of the average charge density, which can be written as the sum of the real part of three complex order parameters $\psi_j(\mathbf{r})$, i.e. $\alpha = \mathrm{Re}\left[\sum_j \psi_j\right]$. The general expression for the Landau free energy density in this case is given by McMillan~\cite{mcmillan1975}:
\begin{equation}\label{eq:mcmillan}\begin{split}
    F = \frac{1}{V}\int d\mathbf{r}\Big[&a(\mathbf{r}) \alpha^2 - b(\mathbf{r}) \alpha^3 +c(\mathbf{r}) \alpha^4 +d(\mathbf{r}) \sum\limits_j |\psi_j\psi_{j+1}|^2 \\ &+e(\mathbf{r})\sum\limits_j|(\mathbf{Q}_j\cdot\nabla - iQ_j^2)\psi_j|^2 +f(\mathbf{r})\sum\limits_j |\mathbf{Q}_j \times \nabla\psi_j|^2 \Big].
\end{split}\end{equation}
The first three terms in this expression are the usual description of the onset order with nonzero values of $\alpha$ through the temperature dependence of the $a(\mathbf{r})$ coefficient. For $1T$-TiSe$_2$, the term proportional to $b(\mathbf{r})$ vanishes because of the lattice symmetry. The term proportional to $d(\mathbf{r})$ describes the interaction between the three CDW components. Finally, the terms proportional to $e(\mathbf{r})$ and $f(\mathbf{r})$ favour the CDW components of the form $\psi_j(\mathbf{r}) = \eta_j e^{i\mathbf{Q}_j\cdot\mathbf{r}+i\theta_j}$, that have propagation vectors of the same length and orientation as the vectors $\mathbf{Q}_j$ at which the electronic susceptibility peaks~\cite{NbSe2_1,NbSe2_2}. 

Given the electronic structure of $1T$-TiSe$_2$ with its strongly peaked susceptibility at $\mathbf{Q}_j = \Gamma L_j$, we can assume fluctuations of the CDW propagation direction to be negligible, and the size of the CDW propagation vector to be constant throughout ordered domains. The order parameter is then of the form  $\psi_j(\mathbf{r}) = \eta_j e^{i\mathbf{Q}_j\cdot\mathbf{r}+i\theta_j(\mathbf{r})}$, with the phase $\theta_j(\mathbf{r})$ constant except across domain walls. Within a domain, the final terms in the free energy can be ignored entirely.

Lattice positions coincide with the maxima of a CDW component whenever its phase $\theta_j(\mathbf{r})$ equals either zero or $\pi$. The interplay of electron-phonon coupling tending to lock individual CDW phases to the lattice and of on-site Coulomb interactions to cause phase variations between CDW components, has been argued to lead to the emergence of orbital order and chiral lattice distortions in the ground state of $1T$-TiSe$_2$~\cite{JvW,MIT,Peng}. It can be described within the Ginzburg-Landau free energy by considering the spatial variations of the coefficients $a(\mathbf{r})$ and $c(\mathbf{r})$. For our present purposes, however, the tiny variations of the phases corresponding to the orbital order are irrelevant, and we assume from here on that  within domains, $\theta_j(\mathbf{r})$ equals zero or $\pi$ for each of the CDW components. 

With these approximations, the free energy density is simplified to:
\begin{equation}\label{eq:Fgen}\begin{split}
    F = \sum_j \Bigg[ \frac{1}{2}a\eta_j^2 
    + \frac{3}{8}c \eta_j^4 + \left( \frac{3}{2} c + d\right)\eta_j^2\eta_{j+1}^2 \Bigg]
\end{split}\end{equation}
From this expression we can directly compare the ground state energies of the CDW phase with only one component having non-zero amplitude (known as a single-$Q$ or $1Q$ CDW), two components being non-zero ($2Q$), or all three contributing (the triple-$Q$ or $3Q$ state).
Writing $\eta_j=\eta$ for the nonzero components, the expressions for the free energy in each of the states become:
\begin{align}
F_{1Q} &= \frac{1}{2}a\eta^2 + \frac{3}{8}c \eta^4 \notag \\
F_{2Q} &= a\eta^2 + \frac{9}{4}c \eta^4 + d \eta^4 \notag \\
F_{3Q} &= \frac{3}{2}a\eta^2 + \frac{45}{8}c \eta^4 + 3d \eta^4.
\end{align}

Expanding around the critical temperature $T_{\text{CDW}}$ and assuming the quadratic coefficient is the only one vanishing at $T_{\text{CDW}}$, all  temperature-dependence is carried by $a(T)$, which we write to lowest order as: $a = \alpha \left(\frac{T-T_{\text{CDW}}}{T_{\text{CDW}}}\right) \equiv \alpha t$. Here $t$ denotes the reduced temperature. For stability, we require the fourth-order terms to be net positive. This imposes restrictions on the coefficients $c$ and $d$ in each phase:
\begin{equation*}\begin{split}
    \mathrm{1Q:}& \hspace{30pt} c > 0 \\
    \mathrm{2Q:}& \hspace{30pt} c > -\frac{4}{9} d \\
    \mathrm{3Q:}& \hspace{30pt} c > -\frac{8}{15} d.
\end{split}\end{equation*}

Finally, minimise the free energies with respect to the order parameter $\eta$ yields $\eta^{\mathrm{min}} = 0$, and correspondingly $F^{\mathrm{min}}=0$ above $T_{\text{CDW}}$ in any of the three phases. Below the critical temperature, where $t<0$, we find: 
\begin{equation*}\begin{array}{lll}
    \mathrm{1Q:} \hspace{20pt}& \eta^{\mathrm{min}} = \sqrt{\frac{-2a}{3c}} & \hspace{20pt} F_{1Q}^{\mathrm{min}} = \frac{-a^2}{6c} \\
    \mathrm{2Q:} \hspace{20pt}& \eta^{\mathrm{min}} = \sqrt{\frac{-2a}{9c + 4d}} & \hspace{20pt} F_{2Q}^{\mathrm{min}} = \frac{-a^2}{9c + 4d} \\
    \mathrm{3Q:} \hspace{20pt}& \eta^{\mathrm{min}} = \sqrt{\frac{-6a}{45c + 24d}} & \hspace{20pt} F_{3Q}^{\mathrm{min}} = \frac{-3a^2}{30c+16d}
\end{array}\end{equation*}

It has been experimentally established that the low-temperature phase of $1T$-TiSe$_2$ is a $3Q$ CDW phase~\cite{DiSalvo}. This means that $F_{3Q}^{\mathrm{min}} < F_{1Q}^{\mathrm{min}}$ and simultaneously $F_{3Q}^{\mathrm{min}} < F_{2Q}^{\mathrm{min}}$. This observation thus imposes the constraint $\frac{d}{c} < -\frac{3}{4}$, which in turn implies: 
\begin{equation} \label{eq:123}
    F_{3Q}^{\mathrm{min}} < F_{1Q}^{\mathrm{min}} < F_{2Q}^{\mathrm{min}}.
\end{equation}
Based on this, we expect that if thermal fluctuations or defects induce any local regions with CDW order other than the $3Q$ ground state, these will primarily be of the $1Q$ type. We will thus ignore the $2Q$ configuration from here on. Notice that fluctuations may also manifest as local $3Q$ regions with shifted phases for one or more of the CDW components, rather than changing to the order to $1Q$. In the absence of defects, phase shifted domains that locally coincide with a ground state configuration are in fact more likely, as any region of $1Q$ order would incur an energetic penalty on top of the cost of forming domain walls.

\subsection{CDW melting}
Long range CDW order requires both the order parameter amplitude $\eta$ to be non-zero throughout a sample, and the phases $\theta_j$ to be coherent over long distances. As a consequence, CDW order can be melted in two qualitatively distinct ways~\cite{McMillan1977}. One possibility has the CDW coherence length remaining long compared to atomic or electronic length scales all the way up to the critical temperature. The CDW order can then be described in terms of a mean field theory for electron-hole pair condensation, which mirrors the BCS theory of superconductivity. As a result, we expect the zero-temperature gap in the electronic structure to be related to the critical temperature by the well-known BCS relation: $2\Delta(T=0) = 3.52 k_{\text{B}} T_{\text{CDW}}$. This behaviour is traditionally known as `weak coupling,' but notice that its defining characteristic is the long coherence length rather than the strength of electron-phonon coupling~\cite{NbSe2_1,NbSe2_2}.

For $1T$-TiSe$_2$, the CDW transition temperature is $T_{\text{CDW}} \approx 200$\,K, while both STS and ARPES data both suggest a gap of around $2\Delta(T\approx 5\,\mathrm{K}) \approx 150$\,meV~\cite{cazzaniga2011,chen2015}. Comparing this to $3.52 k_{\text{B}} T_{\text{CDW}} \approx 60$\,meV, this suggests that $1T$-TiSe$_2$ actually falls in the class of so-called `strong coupling' materials. Again, this regime does not actually require any coupling to be strong, but is rather characterised by a short coherence length. In this case, phonon entropy contributes significantly to the fluctuations mediating CDW melting, and phase fluctuations may be expected to proliferate~\cite{McMillan1977}.

Melting of CDW order through phase fluctuations implies that the order parameter amplitude $\eta$ stays non-zero across the phase transition and even well into the disordered phase. The accumulation of phase fluctuations and the corresponding loss of phase coherence, however, causes the average order to be suppressed, such that $\langle \psi_j \rangle =0$ while $\langle |\psi_j|^2 \rangle >0$~\cite{McMillan1977}. This type of transition has been found in several transition metal dichalcogenides, including for example $2H$-NbSe$_2$~\cite{Borisenko,NbSe2_1,NbSe2_2}. Based on the ratio of gap amplitude and transition temperature in $1T$-TiSe$_2$, we may expect it to also exhibit this type of CDW melting. 

To describe the proliferation of phase fluctuations within the Ginzburg-Landau paradigm, we can follow McMillan and employ a course-grained approximation of the thermally fluctuating CDW state~\cite{McMillan1977}. We thus consider perfectly ordered domains separated by domain walls across which the CDW phase changes. Because the CDW components in $1T$-TiSe$_2$ are all of period two, strong lock-in to the atomic lattice may be expected~\cite{Feng}. We thus assume all phases $\theta_j$ to be either zero or $\pi$ within the ordered domains. Since these values do not allow for topological defects to appear at domain wall crossings, and because the precise shape of the domain is not our primary interest, we furthermore simplify by considering a grid of square domains of size $\xi^2_0$, where $\xi_0$ is the CDW coherence length. The CDW order parameters are constant within each domain, and written as $\psi_j^{mn} = \eta_j \exp(i\theta_{mn})$, with $\theta_{mn} = 0$ or $\pi$, and the indices $m$ and $n$ labelling the in-plane position of the domain. 

To describe the thermal evolution of the domain structure, we first consider the ground state (zero temperature) properties determined by the elastic energy cost of atomic displacements, the electronic energy gain from opening up a gap, the energy cost of locally suppressing the gap at domain walls, and the energy gain/cost of having three simultaneous CDW modes. Fixing the parameter in the free energy expansion to their $T=0$ values, the effects of nonzero temperature can then be included by including the entropy in the description~\cite{McMillan1977}. This is possible because in systems with large $\Delta / k_{\text{B}} T_{\text{CDW}}$ ratio, like $1T$-TiSe$_2$, one may expect the coherence length to be short and the entropy to be dominated by phonon fluctuations. Because the the phonon energies are less than $k_{\text{B}} T_{\text{CDW}}$, their effect can be modelled using classical statistical mechanics. 

With all of these ingredients, the potential (ground state) energy of the system can be denoted as:
\begin{equation*}\begin{split}\label{eq:E}
    E =& \sum_{j,m,n} \Bigg[ \frac{1}{2}a |\psi_j^{mn}|^2 +  \frac{3}{8}c|\psi_j^{mn}|^4 + \left(\frac{3}{2}c + d\right) |\psi_j^{mn}\psi_{j+1}^{mn}|^2 
    +\epsilon \left(|\psi_j^{mn}-\psi_j^{m+1,n}|^2 + |\psi_j^{mn} - \psi_j^{mn+1}|^2 \right)\Bigg]. 
\end{split}\end{equation*}
Here the coefficients $a$, $c$, and $d$ are the same parameters as above, multiplied by an overall prefactor accounting for the integration over the area of a single domain. Because we consider here the potential, ground-state energy, $a$ in this case should be interpreted as the $T=0$ value of the temperature-dependent $a(T)$ used above. The term proportional to $\epsilon$ indicates the energy cost of having domain walls separating regions with different phases. It can be thought of as a discretised or course-grained version of the gradient term proportional to $|\nabla \psi_j(\mathbf{r})|^2$.

Taking the experimental input that the ground state has $3Q$ order, the constraints on the parameters in the free energy that we derived before still hold. Together with the assumption that domain walls cost energy to make, this implies:
\begin{equation*}\label{eq:constraints}
    a < 0 \hspace{30pt} c > 0 \hspace{30pt} -\frac{15}{8}c < d < -\frac{3}{4} c \hspace{30 pt} \epsilon > 0.
\end{equation*}
Here, $\epsilon$ is assumed to be small compared to $a$, such that the overall prefactor of $|\psi^{mn}_j|^2$ is negative, and favours a $3Q$ configuration. 

To determine the mean-field thermal expectation values for the order parameters $\langle \psi_j^{mn} \rangle$ and the fluctuations on top of these, $\delta^j_{mn} = \sqrt{\langle |\psi^j_{mn}|^2 \rangle - \langle \psi^j_{mn} \rangle^2}$, we will start from the simplest possible situation, and then consider more realistic ones. The starting point then, is to first of all assume that the local order parameters $\psi^j_{mn}$ have equal amplitudes for all components and are independent of position. Physically, this corresponds to assuming all domains have local $3Q$ CDW order, and that different domains are sufficiently similar to have approximately the same optimal value of the local order parameter amplitude. Second, we assume the phases of all CDW components simultaneously change by $\pi$ across all domain walls. In reality, the fluctuations should be independent for the three CDW components as well as for different domains. We will return to this oversimplification in the next section. 

Taking the simplified domain distribution, each CDW mode can be considered to exist in the mean-field potential generated by the other modes and the order parameters of neighbouring domains~\cite{McMillan1977}. This results in the following mean-field potential for a single mode:
\begin{equation}\begin{split}
    V_1 =& \frac{1}{2} a |\psi|^2 + \frac{3}{8}c|\psi|^4 + \left( \frac{3}{2} c + d \right)|\psi|^2\langle|\psi|^2\rangle + 2\epsilon\left(|\psi|^2 + \langle|\psi|^2\rangle - 2 \psi \langle \psi \rangle \right).
\end{split}\end{equation}
Using this potential energy, both $\langle \psi \rangle$ and $\langle |\psi|^2 \rangle$ can be solved for self-consistently. Their temperature dependence then arises from interpreting the mean-field averages as (classical) thermal expectation values. That is, we can self-consistently solve the two coupled equations:
\begin{equation}
    \langle \psi \rangle = \frac{\int\limits_{-\infty}^\infty d\psi e^{-V_1/k_{\text{B}}T} \psi}{\int\limits_{-\infty}^\infty d\psi e^{-V_1/k_{\text{B}}T}} \hspace {30pt} \mathrm{and} \hspace{30pt} \langle |\psi|^2 \rangle = \frac{\int\limits_{-\infty}^\infty d\psi e^{-V_1/k_{\text{B}}T} |\psi|^2}{\int\limits_{-\infty}^\infty d\psi e^{-V_1/k_{\text{B}}T}}.
\end{equation}

\begin{figure}[bth]
    \centering
    \includegraphics[width=\linewidth]{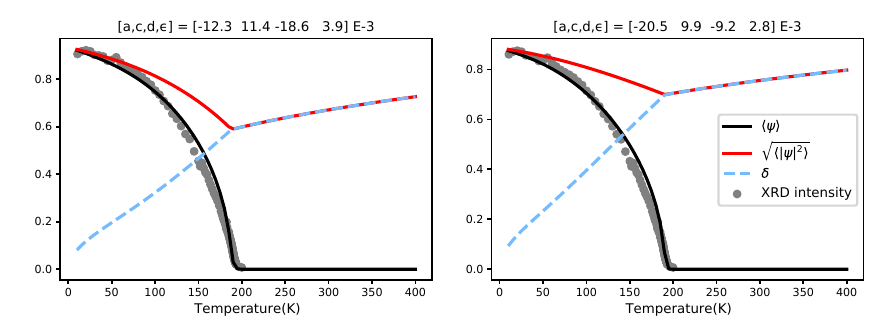}
    \caption{\label{fig:expsi3Q}The thermal expectation values of $\psi$ and $|\psi|^2$ (black and red lines respectively), for two different sets of parameter values, resulting in reasonable qualitative agreement with XRD data from Ref.~\cite{castellan2013} (grey dots). The calculated magnitude of order parameter fluctuations, $\delta = \sqrt{\langle |\psi|^2 \rangle - \langle \psi\rangle^2}$, is shown as a blue dashed line, and tends to zero as temperature decreases.}
\end{figure}

The qualitative shape of the self-consistent solution for $\langle \psi \rangle(T)$ can be compared to the intensity of scattering peaks related to the CDW order, as seen by X-ray diffraction (XRD) experiments. In figure~\ref{fig:expsi3Q}, the temperature-dependent $(1.5,1.5,0.5)$ XRD peak intensity data from Ref.~\cite{castellan2013} is compared to numerical solutions for $\langle \psi \rangle(T)$ given two different sets of parameter values. Although the parameters are constrained somewhat by the relations of Eq.~\ref{eq:constraints}, there is still plenty of freedom within the allowed ranges to obtain reasonable agreement with the experimental data using very different parameter sets. Moreover, the observed peak intensities yield absolute order parameter amplitudes, so that the vertical scale in figure~\ref{fig:expsi3Q} is necessarily in arbitrary units. The comparison of the numerical solution to experimental data should thus be interpreted as a qualitative match, indicating that the model captures the essential physics of the CDW melting. While a more quantitative comparison would require additional experimental constraints, the qualitative agreement between the self-consistent solution and XRD data already gives confidence in the assertion that the melting of CDW order in $1T$-TiSe$_2$ is mediated by phase fluctuations. 

\subsection{Fluctuations}
Having established that the melting of CDW order in $1T$-TiSe$_2$ is likely the result of a proliferation of phase fluctuations, we return to the question what shape these fluctuations are most likely to take. Because of the order of energies in Eq.~\eqref{eq:123}, we know that within any domain, having nonzero amplitudes for all three CDW components is always energetically favourable over having only one or two. This also means that although it is unlikely that the amplitude fluctuations for all CDW components precisely coincide in any domain, large differences between them are still suppressed by exponential Boltzmann factors. This suppression remains intact across the transition temperature, since the average order parameter amplitude remains non-zero (see figure~\ref{fig:expsi3Q}). The assumption of having domains with equal amplitudes $\eta$ for all modes, independent of position, is therefore a reasonable first approximation.

\begin{figure}[bth]
    \centering
    \includegraphics[width=0.8\linewidth]{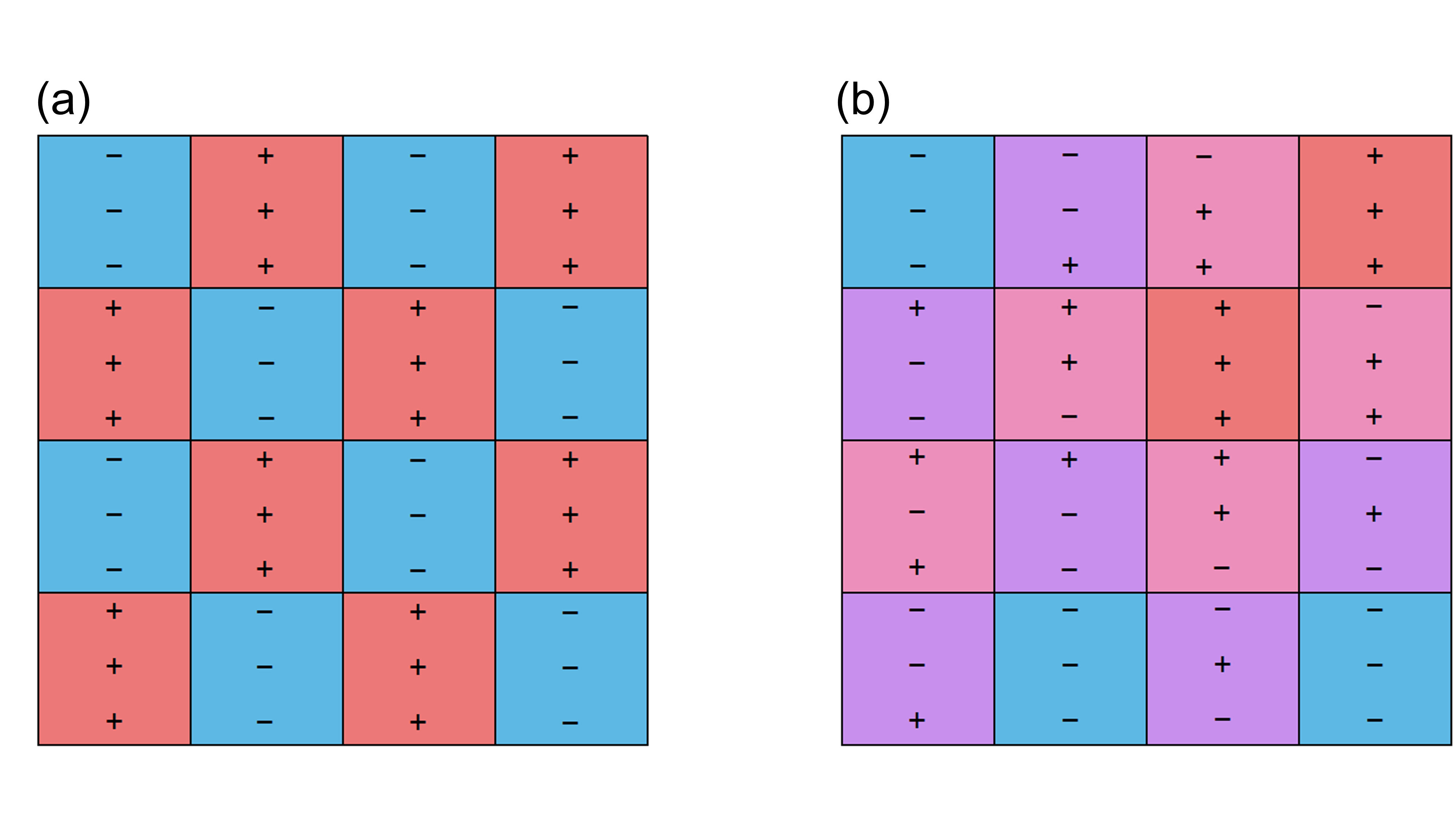}
    \caption{\label{fig:domains}Two possible patterns of phase fluctuation. (a) The pattern used in the mean-field calculations leading to the results in Fig. ~\ref{fig:expsi3Q}. Above the critical temperature, each domain is precisely out of phase with all neighbouring domains in all CDW components. (b) An energetically more favourable configuration, in which only one CDW component switches sign at each domain boundary. Which component switches any particular boundary is randomly selected and does not influence the energy cost of the domain walls.}
\end{figure}

On the other hand, the behaviour of the phases of CDW components across domain walls is determined by the discrete gradient term in the free energy:
\begin{equation}\label{eq:EnergyWall}
    \epsilon \left(|\psi_j^{mn}-\psi_j^{m+1,n}|^2 + |\psi_j^{mn} - \psi_j^{m,n+1}|^2 \right).
\end{equation}
These terms impose an energy cost every time a CDW component changes its phase across a domain wall. In the previous section, we calculated the mean-field values for the order parameter and its fluctuations assuming that all CDW components simultaneously changed their phase at each domain wall. This results in the course grained picture of the fluctuating order shown in figure~\ref{fig:domains}(a). Since Eq.~\eqref{eq:EnergyWall} shows the energy cost of a domain wall increases with every component that changes phase across it, however, a domain structure in which only one of the three components changes sign at each domain wall (as shown in figure~\ref{fig:domains}(b)) will always cost less energy. 

Summing up, the Ginzburg-Landau analysis suggests that above the critical temperature, local fluctuations in the order parameter amplitude may be expected with the same average value for all CDW components. It is always unlikely that any of the three components is significantly weaker than the others, even though CDW modes fluctuate independently. Since neither of the two allowed values for the CDW phase (zero or $\pi$) is energetically favourable within an ordered domain, both may be expected to occur with the same frequency, on average. Neighbouring domains, finally, are likely to differ by the phase of only a single CDW component changing across the domain wall, rather than having all three components undergo a simultaneous phase change.

\subsection{The real space picture}
From the results discussed in the main text, it is clear that the fluctuating domains for the three CDW components all have a longer coherence length in the direction perpendicular to the propagation direction $Q_\text{cdw}$ than parallel to it. This coincides with the description in Ref.~\cite{vanwezel2010}, which argued that one should expect a longer coherence length along one-dimensional ribbons of Ti-Se$_2$-Ti than between adjacent ribbons.

Adding to this experimental observation the qualitative result of the Ginzburg-Landau analysis introduced above, yields a suggestion for the real-space picture of fluctuating domains that may be expected above the critical temperature. In particular, each CDW component may be expected to fluctuate independently from the others, and to anisotropically extend along a different one-dimensional ribbon (perpendicular to its preferred propagation direction) than the others. Three sets of elongated $1Q$ domains, one for each CDW component, are then superposed in the final domain structure to result in a configuration that everywhere has local $3Q$ order, but with domain walls across which only a single CDW component changes phase. The real space cartoon in Fig. 4 of the main text gives an impression of what this may look like.

\section{Structure factors for TiSe$_2$}
\subsection{General formalism}

Elastic X-ray scattering measures the structure factor $S(\mathbf k)$, defined as the squared modulus of the Fourier transform of the scattering potential $S(\mathbf k) = |A(\mathbf k)|^2$ with: 
\begin{equation}
A(\mathbf k) = \frac{1}{2\pi} \int d^3x e^{i \mathbf k \mathbf x} V(\mathbf x). \label{aq}
\end{equation}
If we have a finite system with $N^3$ lattice unit cells labeled by integers $(n_1,n_2,n_3)$ the total scattering potential is given by $V(\mathbf x) = \sum_{n_i}^N V_{n_i}(\mathbf x)$, where $V_{n_i}(\mathbf x)$ is the potential of the $n_i$-th unit cell: $V_{n_i}(x) = \sum_{s,m}f_s\delta(\mathbf x- \mathbf x_{s,m}(n_i))$. Here, $s=$Ti,Se denotes the type of atom in the unit cell, $m$ counts the different atoms of the same species ($m=1$ for Ti, $m=1,2$ for Se) and $\mathbf x_{s,m}(n_i)$ is the position of atom $m$ of species $s$ in unit cell $n_i$. The atomic structure factor for each species is written as $f_s$. The relative value of $f_s$ for Ti and Se can be estimated from the ratio of atomic numbers, $f_{\rm Se}/f_{\rm Ti}=Z_{\rm Se}/Z_{\rm Ti} = 34/22 \approx 1.54$. The scattering potential is therefore $A(\mathbf k) = \sum_{n_i,s,m} f_s e^{i \mathbf k \mathbf x_{s,m}(n_i)}$.

\begin{figure}[tbh]
   \centering
   \includegraphics[width=5.in]{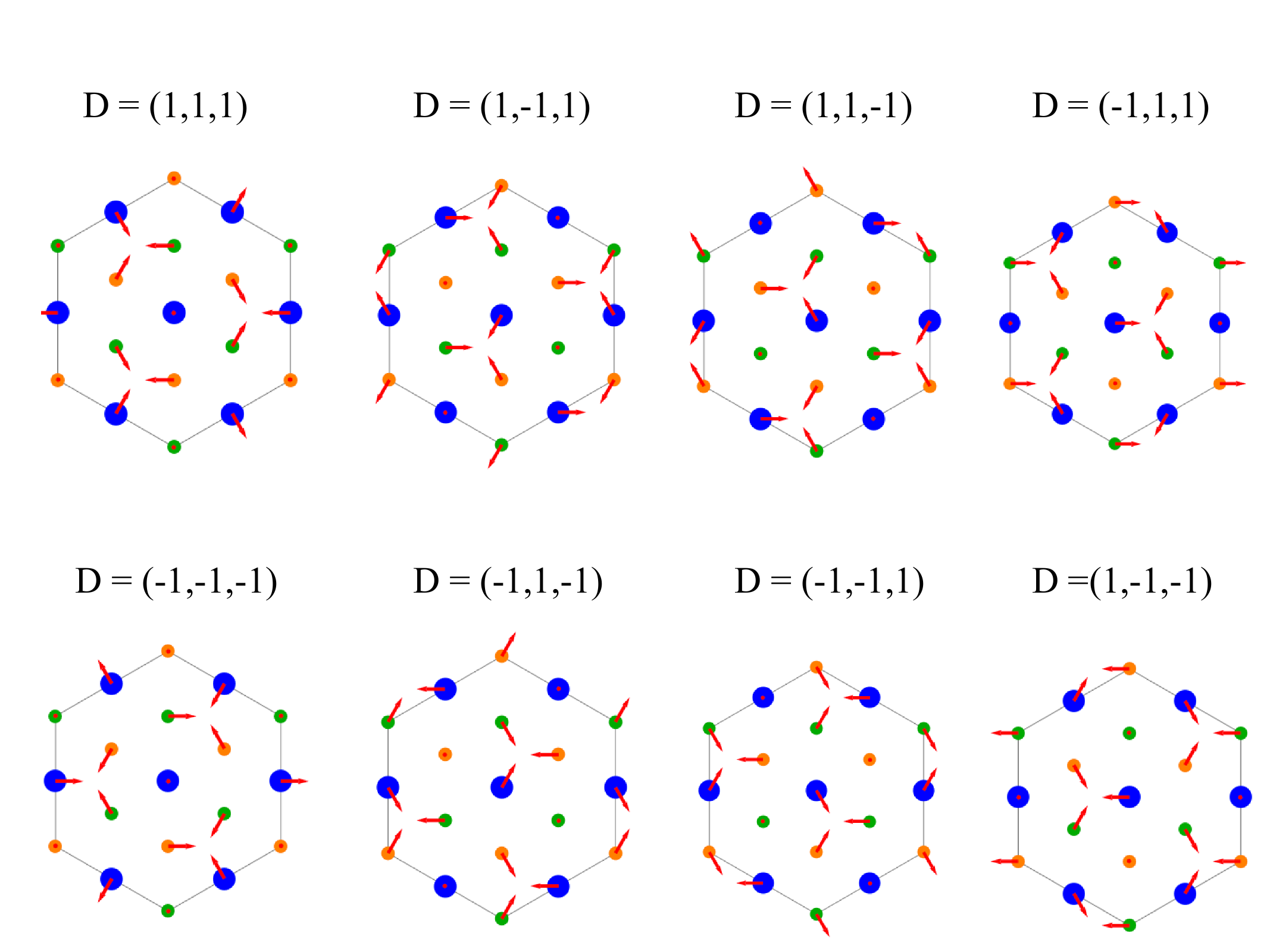} 
   \caption{The 8 possible CDW distortions given by the 8 choices of $D_\alpha$. Note only one layer is shown, while the full pattern contains two layers with the second layer having the opposite displacements.}
   \label{fig:latticeDisplacements}
\end{figure}

In the high-temperature state, $\mathbf x_{s,m}(n_i) = \mathbf x_{s,m}^0 + n_i \mathbf a_i$, where the index $i=1,2,3$ is summed over when repeated, i.e. $n_i \mathbf a_i =  n_1 \mathbf a_1+ n_2 \mathbf a_2+ n_3 \mathbf a_3$, the lattice vectors are $\mathbf a_1 = a(1,0,0)$, $\mathbf a_2 = a(-1/2,\sqrt{3}/2,0)$ and $\mathbf a_3 = c(0,0,1)$ and $\mathbf x_{s,m}^0$ are the high temperature atomic positions in the unit cell. In a general CDW state with domains, these positions are modified to:
\begin{align}
\mathbf x_{s,m}(n_i) = \mathbf x_{s,m}^0 + n_i \mathbf a_i + \sum_\alpha D_\alpha(n_i) \delta \mathbf x_{s,m}^\alpha \cos \mathbf Q_{\rm CDW}^\alpha n_i \mathbf a_i  
\end{align}
Here, $\alpha=1,2,3$ labels the three CDW wavevectors given by $\mathbf Q_{\rm CDW}^1 = (0,\pi/a,\pi/c)$, $\mathbf Q_{\rm CDW}^2 = (-\tfrac{\pi}{2a},\tfrac{\pi\sqrt{3}}{2a},\pi/c)$, $\mathbf Q_{\rm CDW}^3 = (-\tfrac{\pi}{2a},-\tfrac{\pi\sqrt{3}}{2a},\pi/c)$, $\delta \mathbf x_{s,m}^\alpha$ is the associated CDW distortion for each wavevector (See Refs. \cite{DiSalvo76,BCM15}), and $D_\alpha(n_i)$ is a smooth envelope function that allows for domain walls in the order parameter. In our simplified model $D_\alpha(n_i) = \pm 1$ and only sharp domain walls are considered. In the case of constant $D_\alpha$ there are 8 different possible CDW ground states given by the different sign choices in $D_\alpha$, which correspond to the lattice displacements shown in Fig. \ref{fig:latticeDisplacements}.

\subsection{Structure factor in CDW state}

In the high-temperature state where $ \delta \mathbf x_{s,m}^\alpha = 0$ we recover the usual result that the structure factor is only finite for momenta equal to reciprocal lattice vectors $\mathbf G_\alpha = 2 \mathbf Q_{\rm CDW}^\alpha$:
\begin{align}
A_{\rm 1x1x1}(\mathbf k) = \sum_{n_i,s,m} f_s e^{i \mathbf k (\mathbf x_{s,m}^0 + n_i \mathbf a_i)} = N^3 \sum_{s,m} f_s e^{i \mathbf k \mathbf x_{s,m}^0} &, &  \mathbf k = m_1 \mathbf G_1+m_2 \mathbf G_2+m_3 \mathbf G_3
\end{align}

\begin{figure}[tbh]
   \centering
   \includegraphics[width=7in]{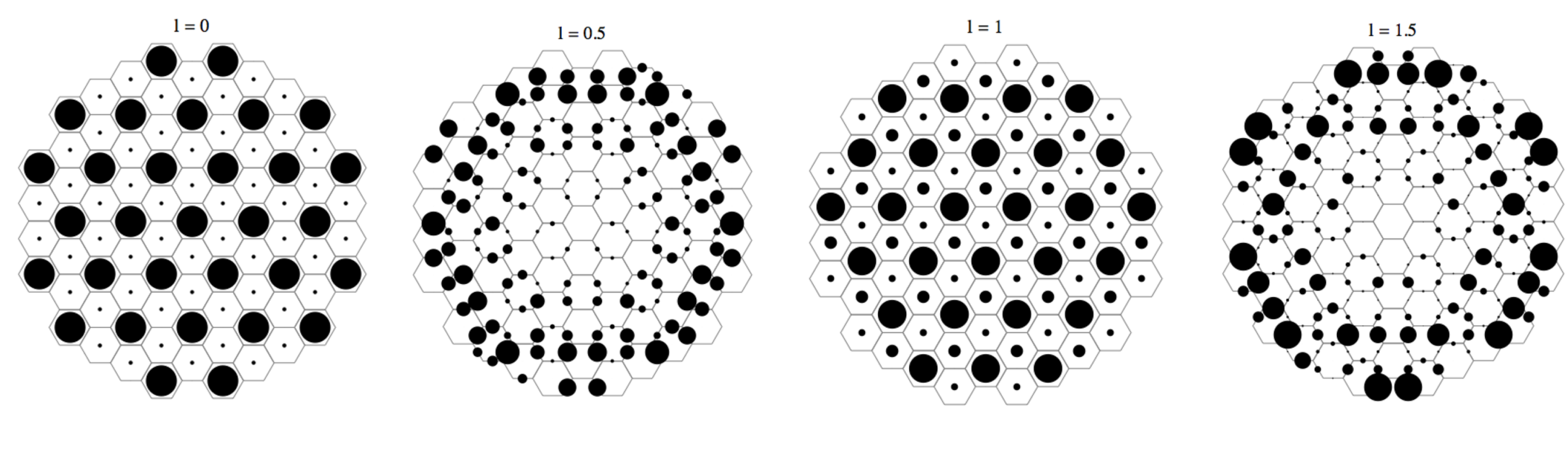} 
   \caption{The structure factor $S(q)$ for different $H-K$ planes. Note that for half integer $l$, $S(q)$ is scaled by a factor ten for ease of comparison.}
   \label{fig:1}
\end{figure}

In the homogeneous CDW state with $D_\alpha=1$, on the other hand, we have:
\begin{align}
A(\mathbf k) &= \sum_{n_i,s,m} f_s \exp (i \mathbf k (\mathbf x_{s,m}^0 + n_i \mathbf a_i + \sum_\alpha \delta \mathbf x_{s,m}^\alpha \cos \mathbf Q_{\rm CDW}^\alpha n_i \mathbf a_i  )) \\ &= (N/2)^3 \sum_{s,m,n_i=0,1} f_s \exp (i \mathbf k (\mathbf x_{s,m}^0 + n_i \mathbf a_i + \sum_\alpha \delta \mathbf x_{s,m}^\alpha \cos \mathbf Q_{\rm CDW}^\alpha n_i \mathbf a_i  )). & & \; \; \mathbf k = m_1 \mathbf G_1/2+m_2 \mathbf G_2/2+m_3 \mathbf G_3/2 \notag
\end{align}
The final sum is evaluated only over a single CDW unit cell $n_i=0,1$ A representation of the structure factor in the CDW state is shown in Fig. \ref{fig:1}, where the area of each disk is proportional to $S(q)$. 

\subsection{Structure factor in a disordered structure}

Now consider a non-periodic structure where the CDW distortion is different for each CDW unit cell, where $D^\alpha(n_i)$ is the domain distribution. The structure factor is given by:
\begin{align}
A(\mathbf k) &= \sum_{n_i,s,m} f_s \exp [i \mathbf k (\mathbf x_{s,m}^0 + n_i \mathbf a_i + \sum_\alpha D_\alpha(n_i)\delta \mathbf x_{s,m}^\alpha \cos \mathbf Q_{\rm CDW}^\alpha n_i \mathbf a_i  )]. \label{full}
\end{align}
This structure factor can only be evaluated numerically, but there is a useful approximation if $|\delta x_{s,m}^\alpha| \ll a,c$ and the argument of the exponential can be Taylor expanded as:
\begin{align}
A(\mathbf k) &= A_{\rm 1x1x1} + \sum_{n_i,s,m} f_s \exp [i \mathbf k (\mathbf x_{s,m}^0 + n_i \mathbf a_i)] (i \mathbf k \sum_\alpha D_\alpha(n_i)\delta \mathbf x_{s,m}^\alpha \cos \mathbf Q_{\rm CDW}^\alpha n_i \mathbf a_i).
\end{align}
We will consider $A(\mathbf k)$ only in the neighborhood of a CDW Bragg peak with wave vector $
\mathbf k = \mathbf Q_{\rm CDW}^\beta +m_i \mathbf G_i + \delta \mathbf k$, 
where $A_{\rm 1x1x1}=0$ and we find:
\begin{align}
A(\mathbf k) &= \tfrac{1}{2}\sum_{n_i,s,m,\alpha} f_s \exp (i \mathbf k \mathbf x_{s,m}^0) [\exp (i n_i (\mathbf k + \mathbf Q_{\rm CDW}^\alpha) \mathbf a_i)+\exp (i n_i (\mathbf k - \mathbf Q_{\rm CDW}^\alpha) \mathbf a_i)] (i \mathbf k  D_\alpha(n_i)\delta \mathbf x_{s,m}^\alpha) \\
& =\tfrac{1}{2}\sum_{s,m,\alpha} f_s \exp (i \mathbf k \mathbf x_{s,m}^0) i \mathbf k \delta \mathbf x_{s,m}^\alpha[D_\alpha(\mathbf k + \mathbf Q_{\rm CDW}^\alpha)+D_\alpha(\mathbf k - \mathbf Q_{\rm CDW}^\alpha)].
\end{align}
The Fourier transform of the domain distribution is here given by $D_\alpha(\mathbf k) = \sum_{n_i} e^{i n_i \mathbf a_i \mathbf k} D_\alpha(n_i)$.
Using the facts that $\mathbf k$ is close to a CDW peak, that $D_\alpha(\mathbf k +m_i \mathbf G_i) = D_\alpha(\mathbf k)$, and that $-\mathbf Q_{\rm CDW}^\alpha = \mathbf Q_{\rm CDW}^\alpha + m_i \mathbf G_i$, we find:
\begin{align}
A(\mathbf k) = \sum_{s,m,\alpha} f_s \exp (i \mathbf k \mathbf x_{s,m}^0) i \mathbf k \delta \mathbf x_{s,m}^\alpha D_\alpha(\delta \mathbf k + \mathbf Q_{\rm CDW}^\alpha+\mathbf Q_{\rm CDW}^\beta).
\end{align}    
If the modulating functions $D_\alpha$ are smooth, they will have small values near large momenta so  $D_2(\delta \mathbf k + \mathbf Q_{\rm CDW}^\alpha+\mathbf Q_{\rm CDW}^\beta) \sim 0$ when $\alpha \neq \beta$, and we obtain:
\begin{align}
A(\mathbf k) &= \sum_{s,m} f_s \exp (i \mathbf k \mathbf x_{s,m}^0) i \mathbf k \delta  \mathbf x_{s,m}^\beta D_\beta(\delta \mathbf k), & 
\notag \\
\Rightarrow ~~~ S(\mathbf k) &= |\sum_{s,m} f_s \exp (i \mathbf k \mathbf x_{s,m}^0) \mathbf k \delta  \mathbf x_{s,m}^\beta|^2 \, |D_\beta(\delta \mathbf k)|^2.  & \mathbf k = \mathbf Q_{\rm CDW}^\beta +m_i \mathbf G_i + \delta \mathbf k \label{SWithDomains}
\end{align} 
This is the main result of this section: the structure factor of a CDW state with domain walls, evaluated near a CDW wavevector, is proportional to the absolute square of the Fourier transform of the domain density. Note that because $\delta \mathbf  x_{s,m}$ is a transverse in-plane displacement, $\mathbf Q_{\rm CDW} \cdot \delta \mathbf  x_{s,m} =0$ for the CDW peaks at $\mathbf Q_{\rm CDW} = (1/2+n,0,1/2)$. However, it is generally finite at any other CDW Bragg peak. 

\subsection{In plane anisotropy of CDW Bragg peaks}

Given the previous discussion, the observed anisotropy in each of the CDW Bragg peaks $\mathbf Q_{\rm CDW}^\alpha$ below $T_{\rm CDW}$ must come from an anisotropic domain distribution of the corresponding CDW component $D_\alpha$. Each $\mathbf Q_{\rm CDW}^\alpha$ peak is an ellipse approximately twice as long in the $\mathbf Q_{\rm CDW}^\alpha$ direction. This suggests that domain walls in the corresponding $D_\alpha$ are longer perpendicular to $\mathbf Q_{\rm CDW}^\alpha$. This makes intuitive sense, as it means that domain walls principally run parallel to wavefronts.

To give an idea of the real space pattern of CDW component domains, we create a simple model with an anisotropy weighting in favor of domain walls perpendicular to $\mathbf Q_{\rm CDW}^\alpha$. It consists of Ising-like variables $D^{\alpha}(\mathbf{x})$ taking values -1 or 1, which correspond to the orientation of CDW component $\alpha$ on a site $\mathbf{x}$ within a hexagonal lattice. We then consider the Hamiltonian:
\begin{align}
H &=-\sum_{\mathbf{x}}\sum_{\alpha=1}^{3}\left\{\sum_{j=1}^{6}\left(J_{\parallel}\mathbf{\delta}_j\cdot\hat{\mathbf{q}}_\alpha+J_{\perp}\left(1-\mathbf{\delta}_j\cdot\hat{\mathbf{q}}_\alpha\right)\right)D^{\alpha}(\mathbf{x}) D^{\alpha}(\mathbf{x}+\mathbb{\delta}_j)+\lambda\Theta^\alpha\left(\mathbf{x}\right)\right\} \notag \\
\text{with}~~~
\Theta^{\alpha}\left(\boldsymbol{x}\right)&=\begin{cases}
\begin{array}{c}
0,\\
1,
\end{array} & \begin{array}{c}
\sum_{j=1}^{6}D^{\alpha}\left(\boldsymbol{x}+\delta_{j}\right)=0\mod 3\\
\text{otherwise}.
\end{array}\end{cases}
\end{align}
Here, $\delta_j$ are the six nearest neighbour vectors  in the hexagonal lattice, and $\hat{\mathbf{q}}_\alpha$ is a unit vector parallel to CDW component $\alpha$.  We choose $J_\parallel=1$ to define a temperature scale, and assume $J_\perp$ is positive (ferromagnetic). The $\lambda$ term favours straight domain walls (and ferromagnetic domains).

A domain distribution is generated by performing a Monte Carlo quench from infinite to zero temperature. The resulting domain sizes are set by the number of time steps after the quench. The image in the main text was generated using $J_\perp/J_\parallel=10$ and $\lambda/J_\parallel=0.25$ with 500 steps.

\begin{figure}[b]
   \centering
   \includegraphics[width=5.in]{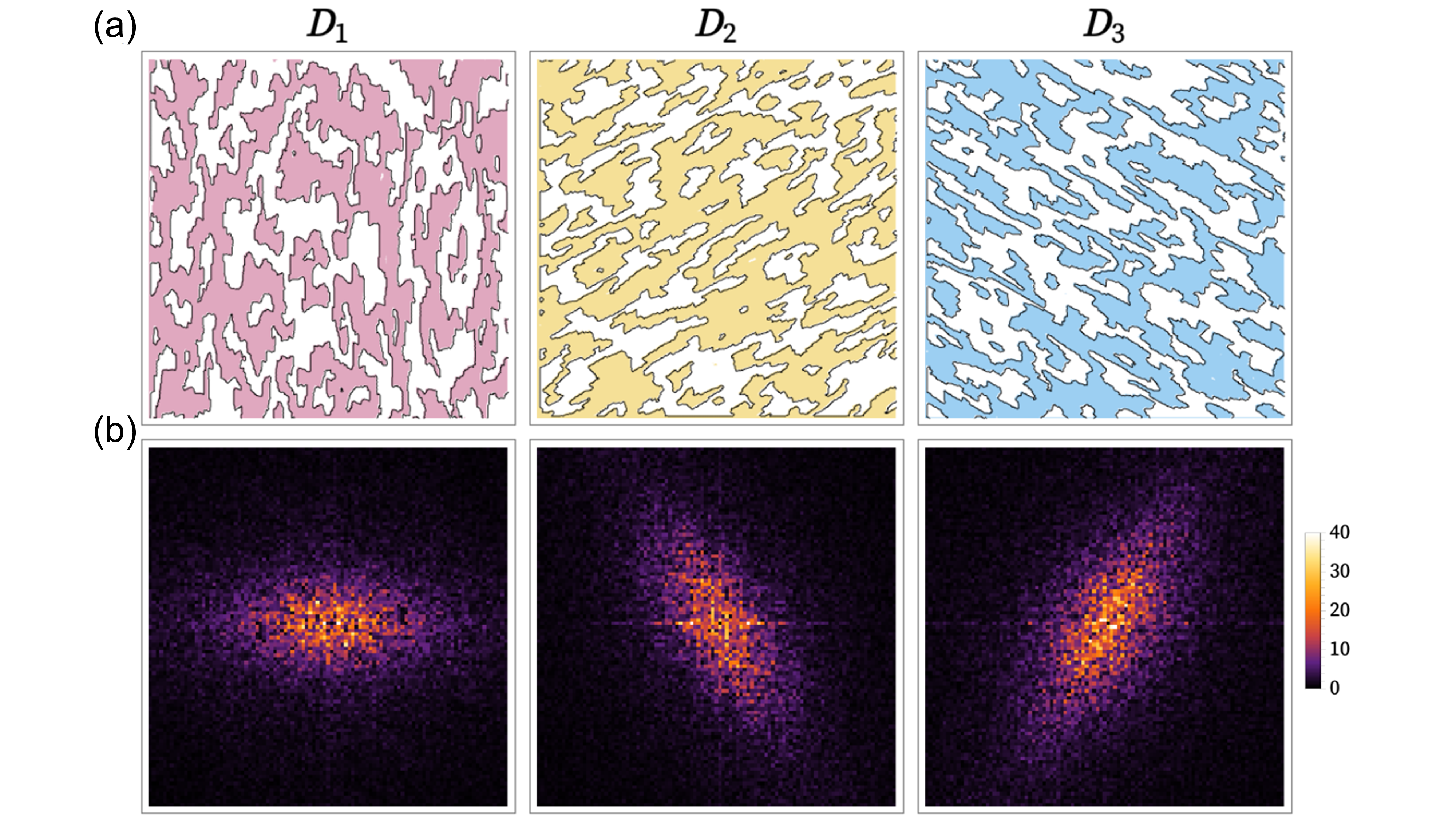} 
   \caption{(a) Domain distributions $D_\alpha$ generated by the Monte Carlo method described in the text. Domain walls are predominantly perpendicular to the corresponding $\mathbf Q_{\rm CDW}$. (b) Fourier transforms $D_\alpha(k_x,k_y)$ of each component.}
   \label{fig:felix}
\end{figure}

The real space domain distribution $D^{\alpha}(\mathbf{x})$ is shown in Fig. \ref{fig:felix}(a). The anistropy in Ising couplings indeed leads to an anisotropic distribution of domains which mostly run perpendicular to $\mathbf Q_{\rm CDW}$. From Eq. \ref{SWithDomains}, the shape of any CDW Bragg peak will be proportional to the Fourier transformed $|D_\alpha(\mathbf k)|^2$, which is shown in in Fig. \ref{fig:felix}(b). The anisotropic domain wall distribution leads to anisotropy in the peak width, as observed experimentally.   

\subsection{Rod structure in $k_z$ from stacking disorder}

To address the rod structure in the $k_z$ direction, we now consider another simplified model where there is full periodicity in the plane, but there is a stacking disorder given by $D_1(n_3)=D_2(n_3)=D_3(n_3)$. To generate the domain distibutions we
assume that if $D_\alpha(n_3)$=1, then $D_\alpha(n_3+1)$ will be 1 with probability $p$ and -1 with probability $1-p$. When $p=0$, we recover the usual $L$-point pattern, and when $p=1$ we recover an $M$-point pattern. For low values of $p$, we obtain a sample with mostly $L$-point character, but a few stacking faults.

\begin{figure}[tbh]
   \centering
   \includegraphics[width=7in]{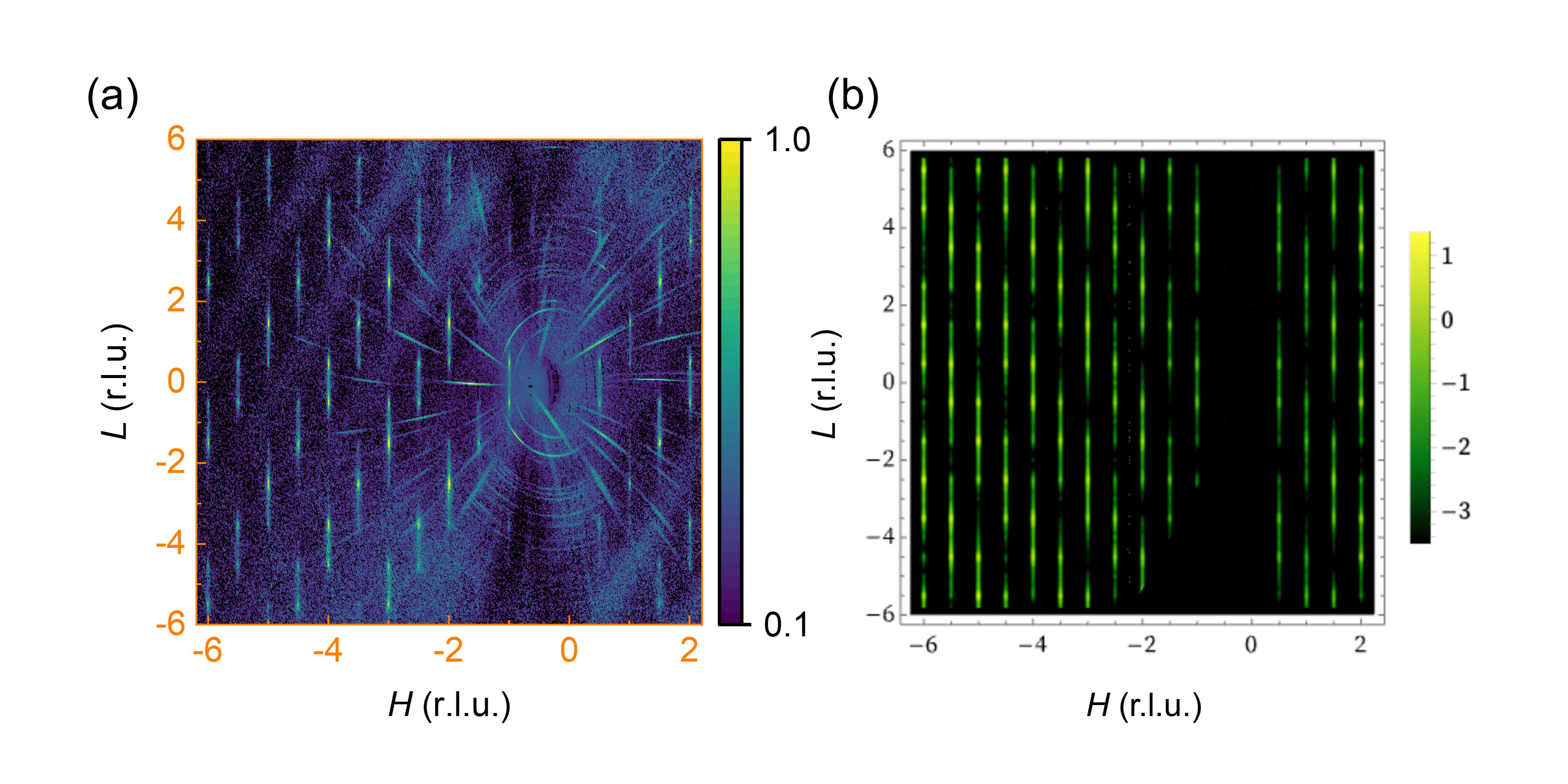} 
   \caption{Structure factor in the $(H, 0.5, L)$ plane. (a) Experimental results at 200 K of the semimetallic sample near $T_{\rm CDW}$. (b) Simulated structure factor with $p=0.1$ giving rise to `rods' similar to those observed in (a).}
   \label{fig:rods}
\end{figure}

Since we aim to describe the whole Fourier space and not just the neighborhood of a Bragg peak, we evaluate Eq. \ref{full} numerically, assuming perfect periodicity in the plane and manually adding an in-plane gaussian broadening to compare with experiments. Using these methods, the structure factor for $p=0.1$ is shown in Fig.~\ref{fig:rods}, next to the experimentally obtained result.

It should be noted that to fix the coordinate system, we need to define the stacking of the Se-Ti-Se layers in the 1$T$ structure. Fig \ref{fig:rods} was generated with $\mathbf x_{Se1}= (a/2,a/2\sqrt{3},-h)$ and $\mathbf x_{Se2}= (a,a/\sqrt{3},+h)$, with $h>0$, while $h<0$ would have produced a reflection of the plot in the $z$ direction. 

\clearpage
\bibliography{suplib}